\documentclass{optica-article}

\journal{opticajournal} 

\articletype{Research Article}
\usepackage{lineno}

\begin{document}

\title{Quantitative Decomposition of Speckle Decorrelation for Inverse Problems in Complex Wave Scattering}

\author{Qihang Zhang,\authormark{1,2} Haoyu Yue,\authormark{1} Ninghe Liu,\authormark{1} Danlin Xu,\authormark{1} Renjie Zhou,\authormark{2} Liangcai Cao\authormark{1} and George Barbastathis\authormark{3,*}}

\address{\authormark{1}State Key Laboratory of Precision Measurement Technology and Instruments, Department of Precision Instruments, Tsinghua University, Beijing, 100084, China\\
\authormark{2}Department of Biomedical Engineering, The Chinese University of Hong Kong, Shatin, Hong Kong SAR, China\\
\authormark{3}Mechanical Engineering, Massachusetts Institute of Technology, 77 Massachusetts Avenue, Cambridge, 02139, MA, USA}

\email{\authormark{*}gbarb@mit.edu} 


\begin{abstract*} 
Optical scattering remains one of the richest phenomena to study in classical optics. Wavefronts that are severely distorted due to scattering still carry significant information about the optical path upstream. Decoding typically starts with the autocorrelation function; yet the precise relationship between it and scatterer statistics still remains unexplored, hindering the model-based method for inverse problems. Here, we reveal that decorrelation due to backward scattering may be quantified as two distinct terms: the first expresses scattering from the surface, whereas the second is due to the volume beneath. The two terms encode higher-order statistics of their respective regimes within the specimen; Experimental studies on representative custom-made scatterers match theoretical predictions with an overall L1-error of less than 0.2$\%$, providing an advanced forward model for various model-based inverse approaches. As a proof-of-concept, we present two examples, scatterer particle size estimation and reconstruction of the incident beam profile, to validate this improvement.
\end{abstract*}
\section*{Main}
Light scattering is induced by inhomogeneous microstructures with disordered spatial distributions of the refractive index. Disordered media distort the phase of the propagating light, resulting in the familiar speckle intensity patterns at the measurement surface: an observer’s retina or a photoelectric transducer \cite{Goodman2015}. Instead of processing the speckle directly, typically it is more productive to use its statistical moments, {\it i.e.} the autocorrelation function, to reduce the huge computational cost. The autocorrelation encodes quantitative information about the statistical structure of both the coherent illumination and scatterer microstructures \cite{Dainty2013,Zhang2023}. While the influence of the illumination has been intensively studied \cite{Bertolotti2012,Katz2014,Shi2022}, the quantitative contribution of scatterers is primarily of interest in this paper.
    
In 1988, Feng et al. identified a cornerstone concept in wave scattering, known as the “memory effect” \cite{Feng1988}. It reveals the statistical correlation in the scattered speckle field by demonstrating that thin scattering layers preserve rotational invariance between incident and output wavefronts within a limited angular range. The existence of the correlation \cite{Haskel18,Zhang21,Banon23} has triggered applications in imaging through scattering media \cite{Bertolotti2012,Li2020, Wei2021, Shi2022,Shi2023, Lu2023, Lee2023} and around corners \cite{Freund1990, Katz2014,Osnabrugge2017, Gigan2022}, 
wavefront shaping \cite{Popoff2010,Mosk2012,Horstmeyer2015,Hsu2017,Jang2018,Yoon2020,Wu2022,Cao2022} and hyper-spectral imaging \cite{Redding2013,Chakrabarti2015,Boniface2019,Wan2021}.
Furthermore, the speckle correlation also plays a key role in characterization of scattering media, such as packages \cite{Pappu2002,Buchanan2005}, foods \cite{Pandiselvam2020}, blood cells \cite{Lee2012}, and powders \cite{Zhang2023, Zhang2024}.  

The amplitude, depth and gradient of refractive index modulation determine the extent of the speckle correlation. Practical applications, such as imaging through turbid media, where decorrelation in thick samples severely limits the field of view \cite{Bertolotti2022,Zhu2022}, demand a fundamental understanding of decorrelation beyond the memory effect. Although an alternative description of such decorrelation effects is in terms of diffuse light \cite{Freund1988,Schott15}, the complexity of multiple scattering in thick media has precluded the development of a comprehensive framework for backscattering to date.

In this article, we propose a superposition model of back-scattered field statistics to address both scatterer thickness and spatial frequency content by employing a first-principle derivation based on the Lippmann-Schwinger equation \cite{Pham2020,Chen2020, Pang2022,Kang2023}. Our result quantitatively decomposes the intensity autocorrelation into two terms: one due to thin-layer effects and one that is volumetric. These two terms exhibit qualitatively distinct behaviours: the first term is consistent with the result from the first-Born approximation in Goodman's book \cite{Goodman2015} and our previous work \cite{Zhang2023,Zhang2024}, evidenced as a slight modulation of the intensity autocorrelation function sidelobes; the second term expresses decorrelation through limiting the sidelobes' visibility. The predictions of the superposition model show good consistency with the experiments, providing a better forward model for various inverse problems\cite{Katz2014,Buchanan2005,Soldevila2023,Osnabrugge2017,Jang2018}. As a proof-of-concept, we demonstrated the improvement induced by the superposition model in model-based particle size estimation and corner imaging from back-scattered light. Both cases make progress simply by substituting the ideal memory model with the superposition model for better accuracy, maintaining the same algorithm and prior information. The physics-induced improvement in our implementations could be compatible with other potential advancements \cite{zhang2026,Zhang2023,Gigan2022,Bertolotti2022} on the algorithmic side, which is beyond the scope of this study.

\begin{figure}[!htb]
\centering
\includegraphics[width=0.6\linewidth]{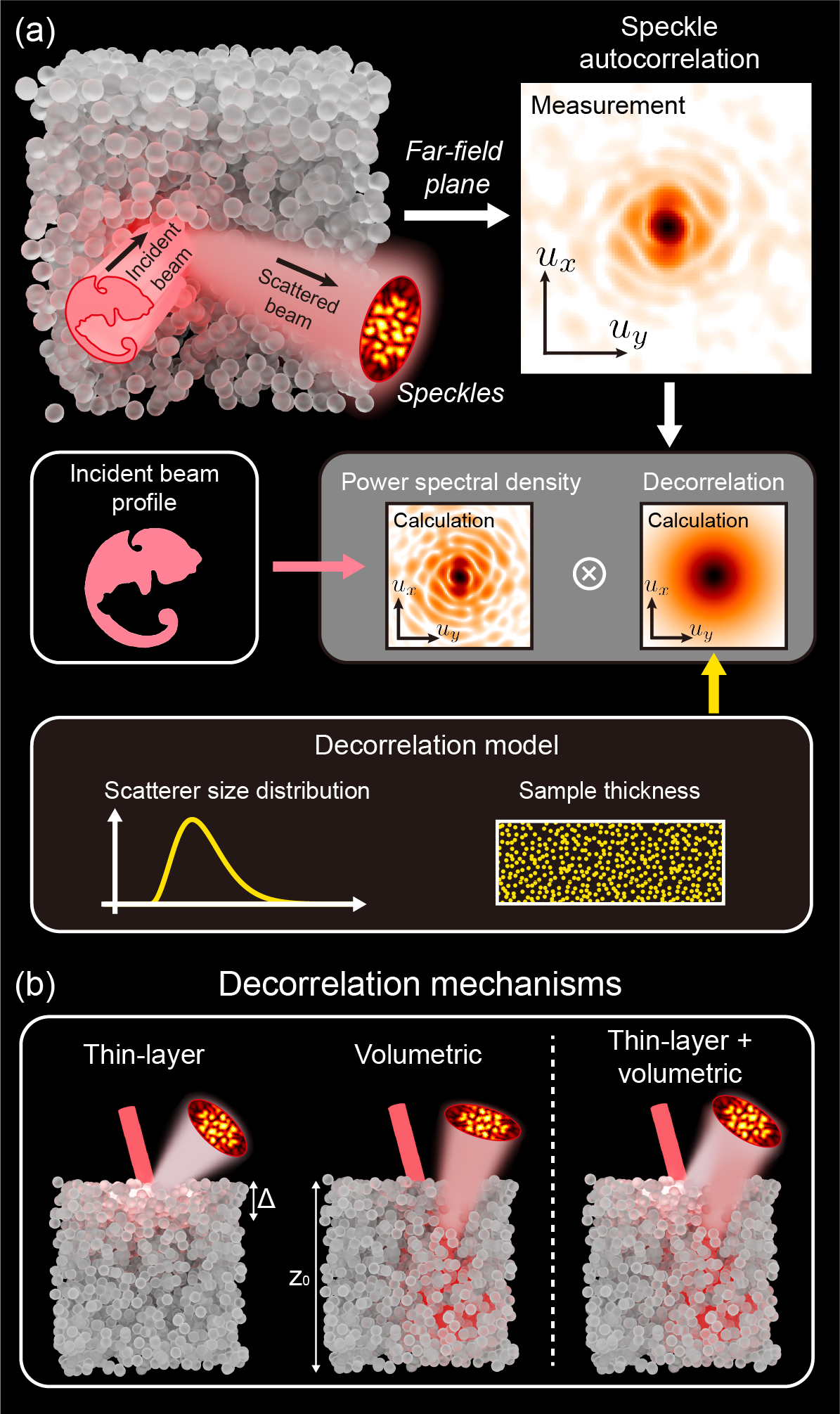}
\caption{ The speckle decorrelation and its decomposition.
(a) Light scattered from a patterned incident beam forms speckle patterns at the far-field plane, whose autocorrelation can be factorized into the beam power spectral density and the scattering decorrelations. For a specific material, sample thicknesses and scatterer size distributions are two main parameters in the decorrelation model. 
(b) A sketch of the decomposition principle in speckle decorrelations. The thin-layer scattering mechanism assumes that the optical field are limited within a top thin-layer $\Delta=330$ $\mu$m, while the volumetric scattering mechanism demonstrates that the scattered light diffuses into the full scatter of thickness $z_0$. For more general scenarios, we reveal that decorrelations can be decomposed into a linear combination of above-mentioned mechanisms.
}
\label{fig:1}
\end{figure}

We assume that the scatterer is of thickness $z_0$ and $p(\pmb{\rho})$ expresses the probability that granular inhomogeneities of size $\pmb{\rho}$ occur. It is illuminated by a spatially patterned coherent beam, as shown in Fig.\ref{fig:1}(a). Let $\mathbf{r}$ denote the Cartesian coordinate in the far field; and let $\psi(\mathbf{r})$ denote the scattered field, assumed to be stationary. The spatial field mutual intensity is, 
\begin{equation}\label{eq:mu_def}
\begin{aligned}
\mu(\mathbf{u}) \equiv \int \psi(\mathbf{r}) \: \psi^{\ast}(\mathbf{r+\mathbf{u}})\text{d}^2\mathbf{r},
\end{aligned}
\end{equation}
where $\mathbf{u}$ is a displacement variable. It can be rewritten as $\mu=\left< \mu \right>+\delta \mu$, where $\left< \mu \right>$ is in the ensemble sense and $\delta \mu$ indicates the zero-mean fluctuation induced by a single realization. According to the Gaussian momentum theorem \cite{Goodman2015}, the ensemble intensity autocorrelation is proportional to $\left<|\mu|^2\right>$, apart from a constant factor \cite{Katz2014}. In a single realization, 
\begin{equation}\label{eq:mu_square_single}
\begin{aligned}
|\mu|^2 = |\left< \mu \right>|^2+|\delta \mu|^2 + 2\text{Re} \{ \left<\mu\right> \cdot \delta \mu^{\ast}\}. 
\end{aligned}
\end{equation}
The last term, $2\text{Re} \{ \left<\mu\right> \cdot \delta \mu^{\ast}\}$, vanishes in the ensemble mean; While the so-called ``shot noise'' term, $|\delta \mu|^2$, is typically far smaller than $|\left< \mu \right>|^2$. We will analyze the ensemble term $\left< \mu \right>$ in the following paragraphs and then return to $\delta \mu$ in the content related to Fig.\ref{fig:4}.

As shown in Fig.\ref{fig:1}(a), it is proved that the intensity autocorrelation, in the other word, $|\left< \mu(\mathbf{u}) \right>|^2$, can be factorized into a product of the pupil's power spectral density and the decorrelation of memory effect for a specific scattering medium, 
\begin{equation}\label{eq:mu_factorization}
\begin{aligned}
|\left< \mu(\mathbf{u}) \right>|^2=\left|\text{FT}\{ I_i \}(\mathbf{u})\right|^2  \left|C\left(\mathbf{u} \right)\right|^2,
\end{aligned}
\end{equation}
where $I_i$ denotes the intensity profile of the illumination beam, $\text{FT}\{ \cdot \}$ is the Fourier transform. $C(\mathbf{u})$ describes the decorrelation between speckle fields from two illuminations with only an angle difference of $\theta=|\mathbf{u}|/f$, where $f$ is the focal length of the lens used to Fourier transform the speckle. The detailed derivations are included in Supplementary Section~2. Based on this principle, we further discovered the decomposition of the decorrelation in the following formula, 
\begin{equation}\label{eq:C_final}
\begin{aligned}
C\left(\mathbf{u} \right) = \frac{1}{1+\kappa} C_{\text{2D}}\left(\mathbf{u};p\right) +\frac{\kappa}{1+\kappa} C_{\text{3D}}\left(\mathbf{u};p,z_0 \right).
\end{aligned}
\end{equation}

\begin{figure}[!htb]
\centering
\includegraphics[width=1\linewidth]{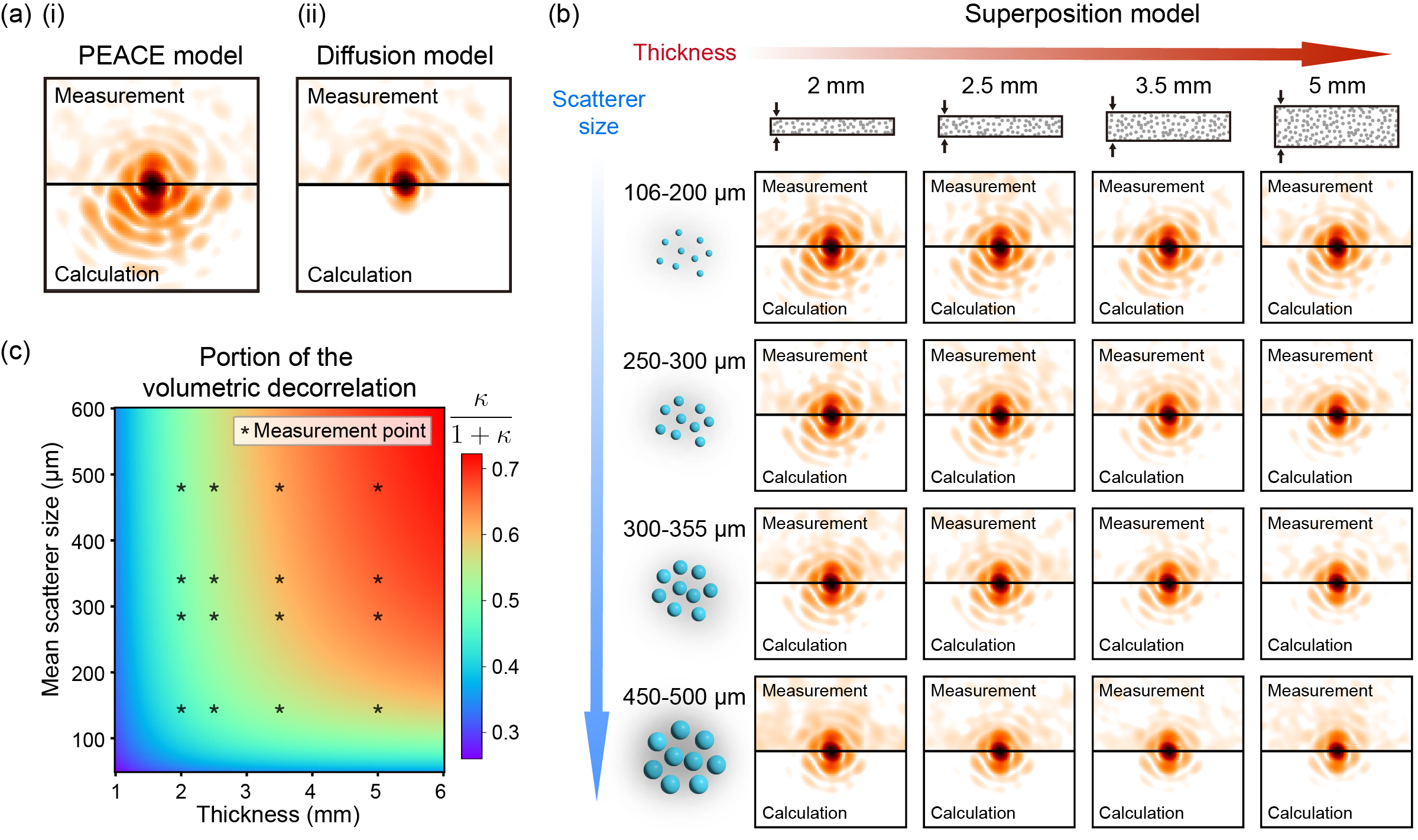}
\caption{ Measurements and fitting results for decorrelations under various conditions. (a) Calculations from both (i) the PEACE model and (ii) the diffusion model fail to match the measurement. The upper panel shows normalized measurements of speckle autocorrelations, ensemble-averaged over 500 images whose scatterer size and thickness are 250-300 $\mu$m and 3.5 mm, respectively. 
(b) Comparison between measurements and fitting results of our superposition model for various thicknesses and scatterer sizes. 
Fluctuations in the measurements come from the residual of finite averaging frames. In (a) and (b), the autocorrelation values are raised to the power of one-fourth to enhance the visibility of sidelobes. 
(c) The color map shows the calculated $\kappa/(1+\kappa)$ coefficient for the diffusion term $C_{\text{3D}}$ as a function of mean scatterer size and sample thickness. Black stars label the corresponding conditions of the measurements in (b).}
\label{fig:2}
\end{figure}

Figure.\ref{fig:1}(b) demonstrates the physical meaning of each term in equation.(\ref{eq:C_final}). 
The surface term $C_{\text{2D}}$ comes from the top ``thin-layer'' region where the single scattering approximation applies.
The approximate thickness of the thin-layer region is $\Delta=\sqrt{\lambda f/2}$, where $\lambda$ is the wavelength. Particularly, $\Delta$ turns out to be independent of the scatterer statistics and determined simply by the Fresnel number $F=\sqrt{\lambda f}/A_p$ as $\Delta = FA_p/\sqrt{2}$, where $A_p$ is the aperture of the imaging system. 
The volumetric term $C_{\text{3D}}$ is similar to the well-known diffusive light \cite{Rossum1999,Cao2022}, which ignores the field interference in an ensemble sense to handle the multi-scattering challenge in the thick volume.
The mixture coefficient $\kappa \equiv \kappa(p,z_0,\Delta)$ depends on scatterer size, sample thickness and approximate thickness of the thin-layer region.
Detailed expressions for $C_{2\text{D}}$, $C_{3\text{D}}$ and $\kappa$ are derived in Supplementary Section~3. These include three unknown parameters to be fitted from data. The parameters relate to absorption, anisotropy and decorrelation depth in the volume. We hypothesize that these unknown parameters are influenced by refractive index statistics but are approximately independent of the scatterer size and the sample thickness. This hypothesis was verified by our experiments described immediately below. 

This decomposition principle is valid under the condition that $z_0 \gg\Delta$. In addition, the lower limit of the spatial frequency component $\pmb{\rho}$ should guarantee that the induced optical path difference $|\pmb{\rho}|\delta n$ is far greater than the wavelength $\lambda$ \cite{Pang2022}, where $\delta n$ is the refractive index difference in scattering media. While the upper limit of $\pmb{\rho}$ should be smaller than the finest feature size in the illumination beam. 

To verify the superposition model in equation.(\ref{eq:C_final}), a Tai-Chi cat shape, as in Fig.\ref{fig:1}(a), was used as incident beam profile. This was chosen because its power spectral density contains continuous low-frequency sidelobes along all directions as well as sufficient high-frequency components. Experimental details and data for more incident patterns are plotted in Supplementary Section~1 and 4, respectively. As scattering media, calibrated  of potassium chloride (KCl) powders of four scatterer size distributions $p(\pmb{\rho})$ and four thicknesses $z_0$ were prepared, for a total of 16 data sets. Figure.\ref{fig:2}(a) shows that both PEACE \cite{Zhang2023} and pure diffusion models fail to match the measured intensity correlation functions well. The PEACE model claims little decorrelation while the diffusion model predicts a decorrelation much stronger than the measurement. Instead, in Fig.\ref{fig:2}(b) our superposition model is shown to balance well the contributions of thin-layer and volumetric scattering, offering a better match to the experimental results. The three parameters fit the 16 different experimental images with an overall L1-loss of less than $0.002$. Details of the fitted parameters are in Supplementary Section~4. 
Figure.\ref{fig:2}(c) evaluates the relative contribution $\kappa / (1+\kappa)$ due to the diffusion term in the superposition model. It is observed that the thin-layer contribution $C_{\text{2D}}$ dominates the decorrelation in either thin samples or small scatterers. 
By contrast, for thick samples and large scatterers at the top-right corner, volumetric decorrelation $C_{\text{3D}}$ becomes dominant. 

\begin{figure}[!htb]
\centering
\includegraphics[width=0.7\linewidth]{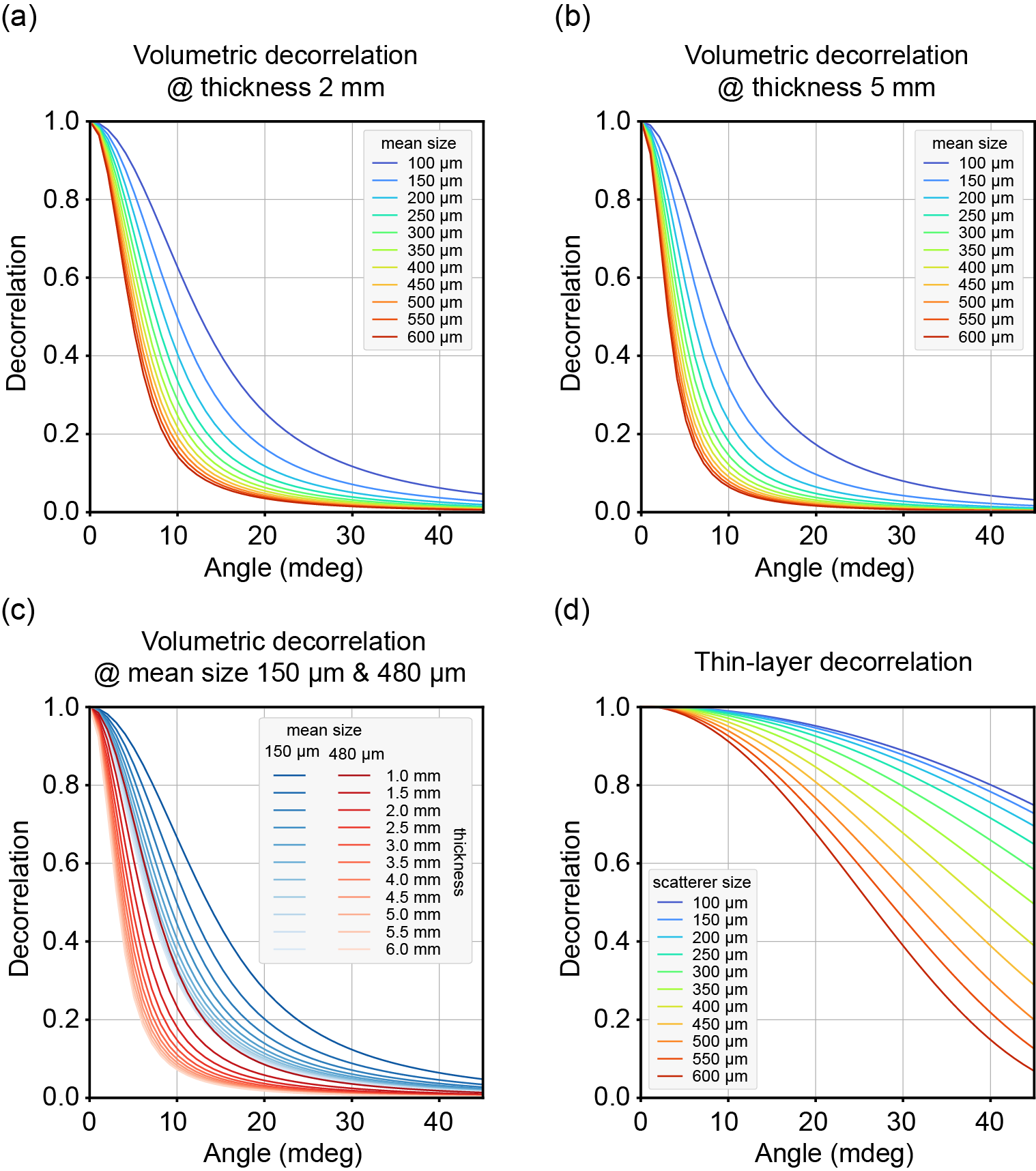}
\caption{ Sample thickness and scatterer size dependence of $C_{\text{2D}}$ and $C_{\text{3D}}$.
(a) Computed $C_{\text{3D}}$ curves for varying mean scatterer sizes at a thickness of 2 mm. 
(b) Computed size-dependent curves of $C_{\text{3D}}$ at a thickness of 5 mm. 
(c) Calculated $C_{\text{3D}}$ for different sample thicknesses at mean scatterer sizes of 150 $\mu$m (blue curves) and 480 $\mu$m (red curves). 
(d) Calculated $C_{\text{2D}}$ for different scatterer sizes. The calculation in this plot presumes the scatterer size distribution to be a $\delta$-function centered at a specific size. The x-axis in (a)-(d) is the displaced angle $\theta =|\mathbf{u}|/f$.
}
\label{fig:3}
\end{figure}
%
Figure.\ref{fig:3} assesses the angular behavior of the volumetric scatter {\it vis-\`{a}-vis} the thin layer effect. Figure.\ref{fig:3}(a) and (b) present the volumetric decorrelation for sample thicknesses $z_0=2$ mm and $5$ mm, respectively; whereas Fig.\ref{fig:3}(c) is for mean scatterer sizes $\left<|\pmb{\rho}|\right>=150$ $\mu$m and 480 $\mu$m. Figure.\ref{fig:3}(d) is thin-layer decorrelation for different mean particle sizes. These results reveal several qualitative differences between the two scattering mechanisms. {\it (i)} Volumetric scattering decorrelates faster than the thin-layer term under the same conditions. {\it (ii)} The second derivatives of the two decorrelation mechanisms exhibit opposite signs in the angle range of 10 - 35 mdeg, with the volumetric decorrelation being concave and the thin-layer decorrelation being convex. {\it (iii)} Within the plotted size range, volumetric decorrelation curves are more sensitive to scatterer size for smaller scatterers, and saturate at larger scatterer sizes. Conversely, the sensitivity of the thin-layer decorrelation reaches its minimum at the small-size limit.
Under the extreme condition $z_0 \gg |\pmb{\rho}|$, the effective width of $C_{\text{3D}}$, which is plotted in Supplementary Fig.S5, follows the power law \cite{Feng1988} and is proportional to $\lambda f/l$, where the mean free path $l$ is equivalent to the mean scatterer size $\left<|\pmb{\rho}|\right>$ in this paper.

\begin{figure}[htb!]
    \centering
    \includegraphics[width=0.7\linewidth]{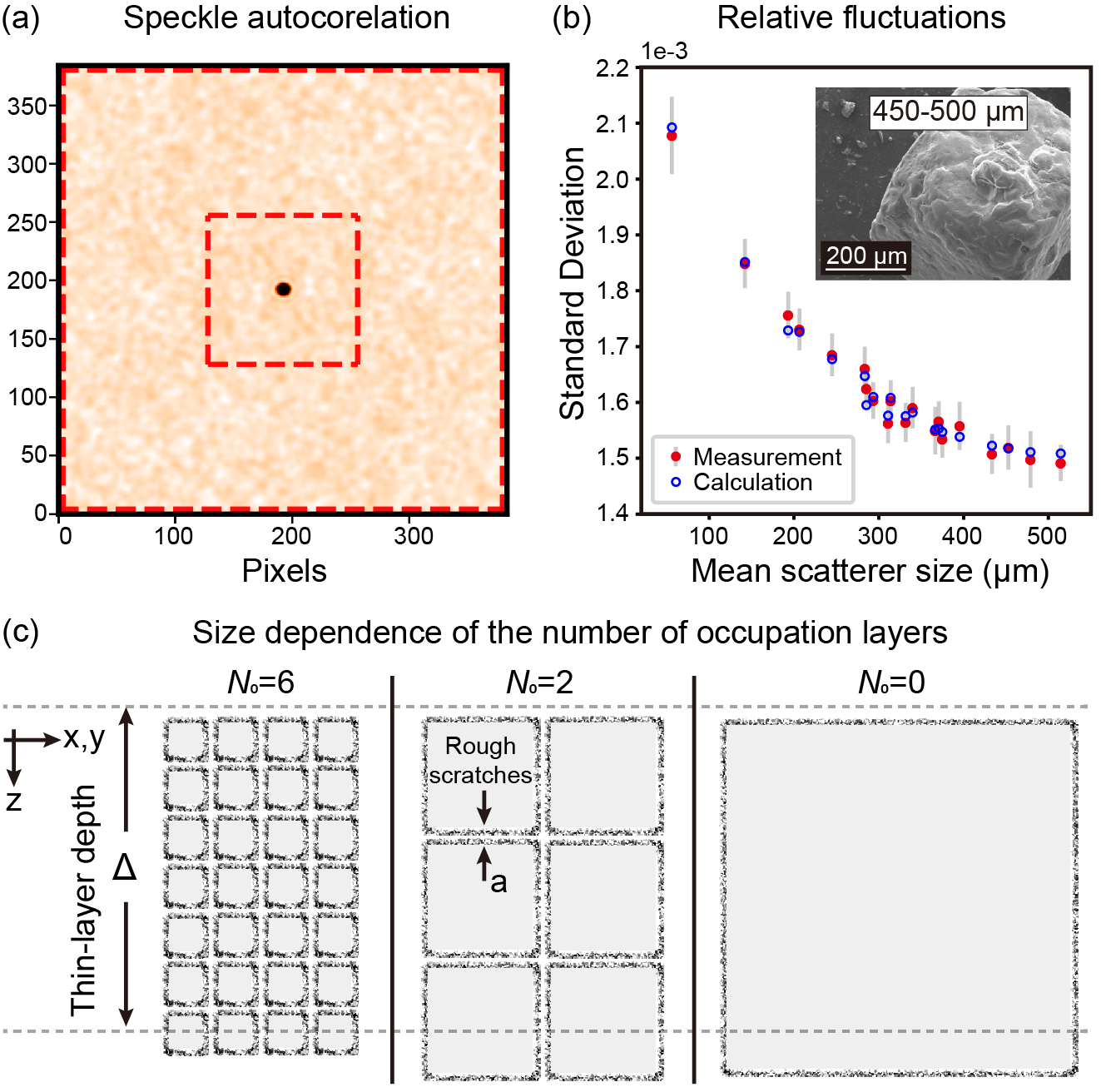}
    \caption{ A model for the fluctuation in a single realization. (a) A single realization example of the speckle autocorrelation. Fluctuations at the large displacement can be observed, which is characterized by the standard deviation in the marked region between two red dashed squares. 
    (b) Fluctuations for 20 size distributions as a function of mean scatterer size. 
    Each red measurement dot and gray error bar represent the ensemble mean and the standard deviation of 500 fluctuations collected from the same sample at different spatial positions, respectively.
    Blue circles indicate the fluctuation values fitted by equation.(\ref{eq:bcg}), given the scatterer size distributions $p(\pmb{\rho})$. Inset: SEM images for KCl powders at $\times$ 180 magnification.
    (c) Physical picture of the fluctuation as a function of scatterer size. Within the thin-layer depth $\Delta$, the number of rough-scratch layers $N_0$ varies with the scatterer size. $a$ is the depth of the rough layers for each scatterer, assumed to be independent of scatterer size. Smaller scatterers can allocate more rough surfaces into the thin-layer region. While the scatterer size is larger than $\Delta$, only one scratch surface stays in the top thin-layer region and the size dependence vanished. 
    }
    \label{fig:4}
\end{figure}

Next, we come back to analyze the fluctuation $\delta\mu$ in a single realization. Goodman's textbook \cite{Goodman2015} denoted  $|\delta \mu|^2$ as shot noise in surface scattering ($z_0 < \Delta$), where the noise level increases with the lateral particle size. However, in our interested scope, $z_0 \gg \Delta$, a counterintuitive tendency is observed that the fluctuation $|\delta\mu|^2$ decreases with the size of the scatterer. 
These relative fluctuations in single realizations were quantified by the standard deviation among the extended region, marked by red dashed lines in an example displayed in Fig.\ref{fig:4}(a). Their ensemble result, $\sqrt{\left<|\delta\mu|^2\right>}/\left< \mu \right>$, are plotted in Fig.\ref{fig:4}(b), for 20 size distributions with a sample thickness of $z_0=1$ cm. According to the detailed derivations in Supplementary Section~5, we obtain
\begin{equation}\label{eq:bcg}
\begin{aligned}
\frac{\sqrt{\left<|\delta\mu|^2\right>}}{\left< \mu \right>} = c + \frac{a^2}{\Delta^2} \int \left[ \frac{\Delta}{|\pmb{\rho}|} \right] p_v(\pmb{\rho})\text{d}\pmb{\rho},
\end{aligned}
\end{equation} 
where $p_v(\pmb{\rho})=p(\pmb{\rho})|\pmb{\rho}|^3 / (\int p(\pmb{\rho})|\pmb{\rho}|^3 \text{d}\pmb{\rho})$ is the volume-based scatterer size distribution. The constant $c$ results from volumetric scattering deeply inside the sample and is approximately independent to scatterer size. The rest term in equation.(\ref{eq:bcg}) is proportional to the volume portion of rough scratches on powder particle surfaces within the thin-layer region.
A scanning electron microscopy (SEM) image for the KCl cubic crystal particle in Fig.\ref{fig:4}(b) inset suggests that the depth of random scratches $a$ could be assumed to be independent to the scatterer size for simplicity. More SEM images are included in Supplementary Fig.S8.
Because powder particles are densely located, the lateral area of scratches is irrelevant to the scatterer size. Thus, its volume portion is mainly determined by the number of occupation layers $N_0$ along the z-direction within an approximate depth of $\Delta$. 
As represented in Fig.\ref{fig:4}(c), the size dependence of $N_0$ is $N_0 = [\Delta/|\pmb{\rho}|]$, where $[\cdot]$ is the floor function. 
The overall contribution of thin-layer scattering is proportional to $N_0$ with a coefficient of $a^2/\Delta^2$.
Parameters $a$, $\Delta$ and $c$ were fitted by 20 samples with different distributions, all fitted points align well with the corresponding measurements, as plotted in Fig.\ref{fig:4}(b).
The fitted parameters $a_{\text{fit}}$, $c_{\text{fit}}$ is $1.80$ $\mu\text{m}$ and $2.95\times 10^{-3}$, respectively. The fitted thickness $\Delta_{\text{fit}}=316$ $\mu$m closely approximates the calculated value 330 $\mu$m in our system, reinforcing the validity of the physical picture in Fig.\ref{fig:4}(c). 
Due to the floor function, particles with a size $|\pmb{\rho}|>\Delta$ have no contribution to equation.(\ref{eq:bcg}), which is consistent with the observation in Fig.\ref{fig:4}(b) that the fluctuation gradually loses the size dependence for the mean scatterer size above 400 $\mu$m.
Equation.(\ref{eq:bcg}) suggests that the fluctuation of mutual intensity in a single realization $\delta \mu$ is not pure noise, as it encodes the scatterer information explicitly.

\begin{figure}[!htb]
\centering
\includegraphics[width=0.7\linewidth]{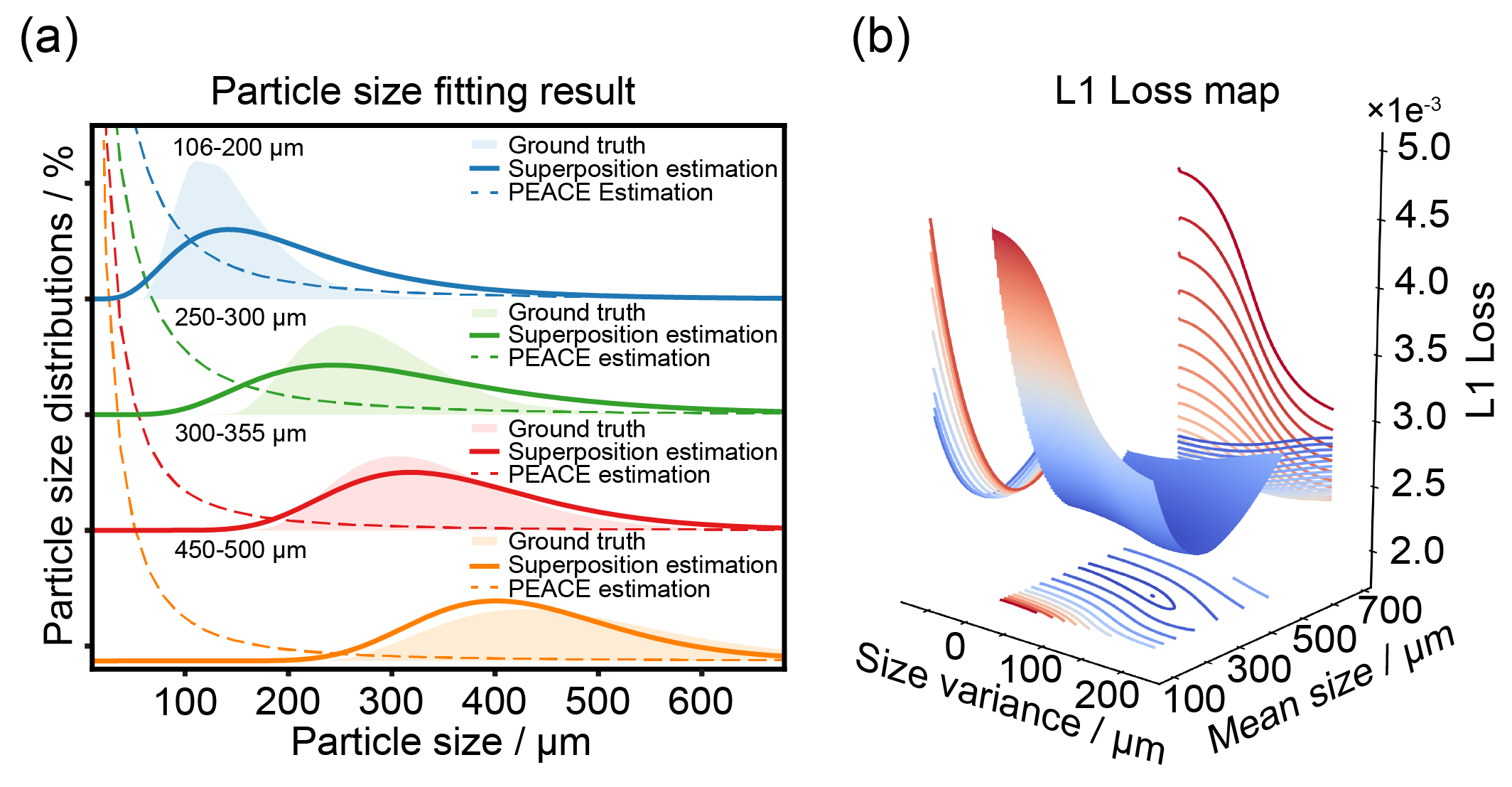}
\caption{Model-based Particle Size Estimation. (a) With a given incident beam profile, fitting results of the particle size distribution by the PEACE model and the superposition model are presented. The ground truth is calibrated by a commercial offline particle size analyzer, the “Mastersizer 3000”. Size estimations from the superposition model match the ground truth for the mean particle size and have slight deviations for the width. The PEACE model fails to predict correct size distributions due to the incompleteness of the forward model. (b) The L1-loss is a function of the mean size and the size variance in the lognormal distribution. The map is plotted for the sample size of 250-300 $\mu$m. Projections along different directions indicate that the fitted result is sensitive to the mean size but insensitive to the size variance, accounting for the mismatch in the width of the fitted distributions. 
}
\label{fig:5}
\end{figure}

In Fig.\ref{fig:5}, we demonstrate the model-based particle size estimation, given the intensity of the incident beam $I_i$. Assuming that particle sizes $\rho$ follow a unimodal Lognormal distribution $p(\rho)$ and opting for the simplest enumeration method, we plotted an L1 loss map between the calculated result of equation.(\ref{eq:C_final}) and the measurement, as a function of mean size and size variance shown in Fig.\ref{fig:5}b. Upon identifying the minimum point, we got the optimal distributions, as plotted together with the ground truth in Fig.\ref{fig:5}a. The mean value of the fitted distribution aligns with the ground truth, while the width is consistently slightly broader. The projections in Fig.\ref{fig:5}b demonstrate that the L1-Loss is sensitive to the mean size but insensitive to the variance in size , which accounts for the inaccurate estimation of the distribution width. As illustrated by the dashed line in Fig.\ref{fig:5}a, fitting results based on the PEACE model significantly deviate from the ground truth. Although learning-based estimation provides a faster and more accurate prediction\cite{Zhang2023}, this model-based estimation offers superior stability and generalization ability. Moreover, the model-based estimation also holds the potential to incorporate machine learning algorithms and amalgamate the advantages of both methods, though the specifics of this strategy are beyond the scope of this work. 
    
 Finally, we perform the reconstruction of the incident beam intensity $I_i$ from the back-scattered speckle pattern, given a known particle size distribution $p(\rho)$. According to equation.(\ref{eq:mu_factorization}), we could reconstruct the incident beam from speckle’s autocorrelations, with different decorrelation models. Previous attempts based on the memory effect, have been made\cite{Katz2014}. Here, We only switched the forward model from the ideal memory model to the superposition model in equation.(\ref{eq:mu_factorization}) and applied the same Gerchberg–Saxton algorithm for image reconstruction. Further details are available in the Supplementary Section~6. Figure.\ref{fig:6}a presents the reconstruction results: The first column is binary intensity pupils served as objects; The second column is the corresponding power spectral density while the third column shows measured speckle autocorrelations, the decorrelation term suppresses sidelobes of the pupil power spectral intensity; Reconstructions from both forward models are presented in columns (iv) and (v). As the decorrelation term is quantitatively compensated in the phase retrieval algorithm, the reconstruction from the superposition model contains more high frequency features, resulting in a better image quality. Figure.\ref{fig:6}b shows a series of reconstructed images with varying powder particle size and thickness for different models, whose ground truth is the cat image in Fig.\ref{fig:1}a. Ideal memory effect can conduct a good image reconstruction for the 2 mm thin sample with the smallest particle size 106-200 $\mu$m, but its image quality drastically decays for thick samples and large particles due to the decorrelation of the shift-invariance. In contrast, reconstructions from the superposition model are relatively stable among different thicknesses and particle sizes.

\begin{figure}[!htb]
\centering
\includegraphics[width=0.7\linewidth]{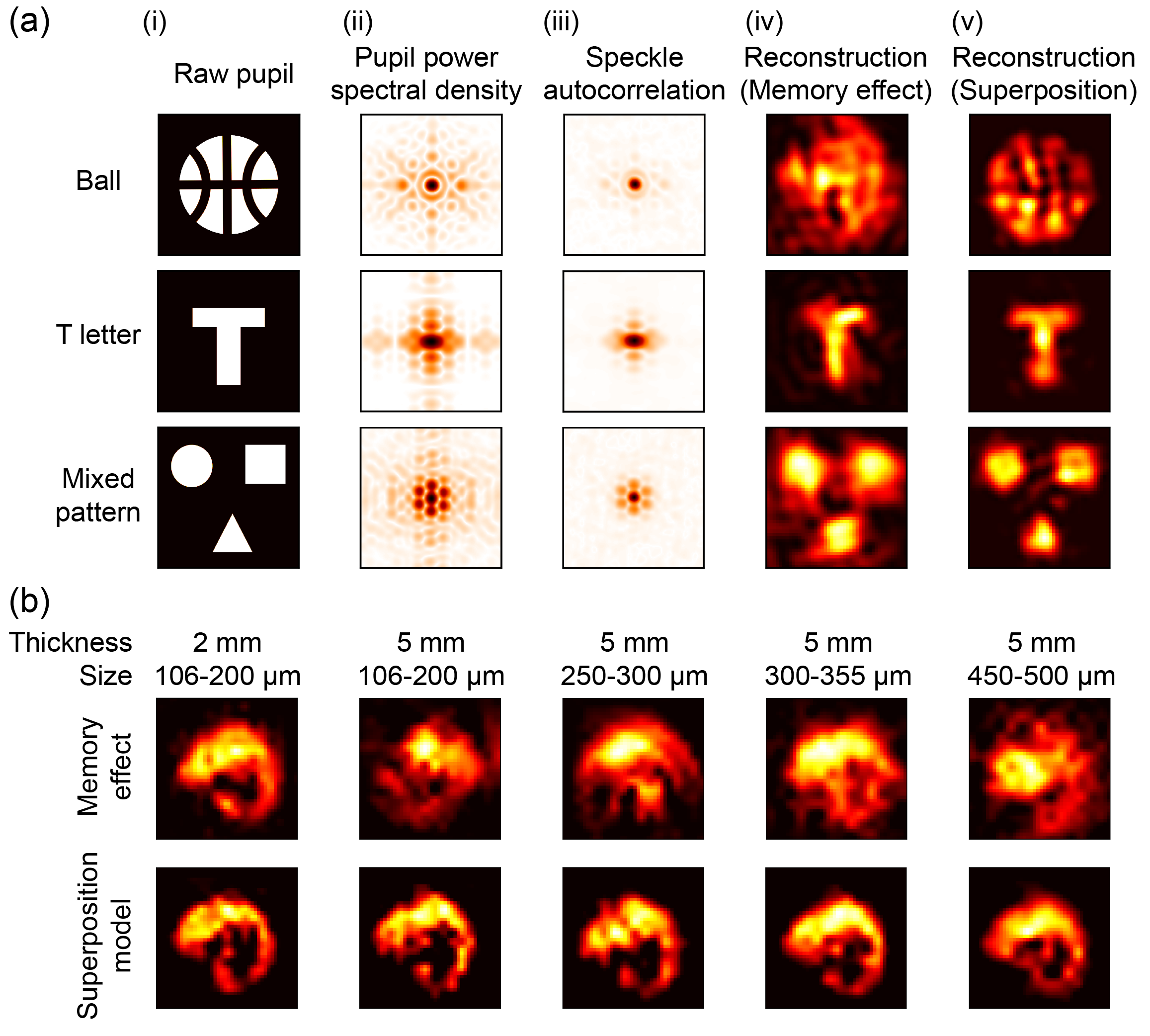}
\caption{ Reconstruction of the incident intensity profile. (a) (i) Patterned pupils to reshape the incident beam profile; (ii) The corresponding power spectral densities of pupils; (iii) Speckle autocorrelations measured from a 250-300 $\mu$m size and 2 mm thickness powder sample; (iv) Reconstructions from the ideal memory model without decorrelation; (v) Reconstructions with quantified decorrelation, the superposition model. (b) The reconstructed image quality evolution with powder particle size and thickness. The ground truth is the cat image in Fig.\ref{fig:1}a. The upper row shows images reconstructed from the ideal memory effect, while the lower panel shows results from the superposition model. 
}
\label{fig:6}
\end{figure}


 
In conclusion, we have identified a fundamental, yet previously missing, principle for the statistics in complex wave scattering, the decomposition principle, which allows for a comprehensive and general description of speckle decorrelations. 
Similarly to the ``emergence" phenomena, a large number of scattering events produce a new simple law, which is different from the one for a single scatterer. Although speckle patterns originate from a complicate multi-scattering process inside the medium, the decomposition principle claims that its autocorrelation is independent of the specific microstructure of scatterers and the corresponding back-scattered field. For forward scattering, the decorrelation only comes from volumetric scattering, as all transmitted light has to penetrate the full depth. The forward decorrelation would deviate from the diffusion model, and the transmission matrix is commonly used to quantify its scattering property, though in this paper we are not investigating this regime any further.  
Moreover, the decorrelation depth along the z-direction remains to be discussed. Here we simply assume that the decorrelation depth is longer than the thin-layer thickness at the sample surface, and it becomes approximately a constant deep inside the scattering volume. In practice, it should be a continuous function of depth, whose specific formula requires further studies. 
Two inverse problems, particle size estimation and corner imaging, show better performances by substituting the ideal memory model with the superposition model. For any prior knowledge and inverse algorithm, it's possible to gain a better performance by upgrading the forward model. Consequently, we anticipate that the amnesia effect could also enhance inverse problems in other scattering related fields\cite{Soldevila2023,Cao2022,Yoon2020,Faccio2020,Zhang2023,Jayabarathi2022,Pandya2023,Li2021,Liu2022}, without necessitating significant changes to their existing algorithms.
\begin{backmatter}

\bmsection{Acknowledgment}
Authors acknowledge S. Mondal, B. Tan, L. Han and S. Dong for their useful discussions and suggestions.
G. B. acknowledges the financial support from Millennium Pharmaceuticals, Inc. (a subsidiary of Takeda Pharmaceuticals) D824-MT15 and the United States Air Force Office of Scientific Research (AFOSR) through the Multi-University Research Initiative (MURI) program. Q. Z. acknowledges the financial supported from the National Natural Science Foundation of China (62405156).
    
\bmsection{Disclosures}
The authors declare no conflicts of interest.

\bmsection{Data Availability Statement} 
The datasets and codes, generated and analyzed during the current study, are available from the corresponding author on reasonable request. 

\bmsection{Authors' Contributions}

G.B. and Q.Z. conceived the project and developed the theory. 
    Q.Z. and H.Y. designed the experiment and collected the data.
    Q.Z. and N.L. developed the code and processed the data.
    H.Y. and D.X. collected the electron microscopy images. 
    R.Z. pointed out the applicable range of this framework. 
    Q.Z., L.C. and G.B. prepared the original manuscript. 
    R.Z., H.Y., N.L. also contributed to the manuscript. 
    G.B. and L.C. supervised this project.

\bmsection{Supplemental document}
Supplementary.pdf contains methods, sample preparations, equation derivations, and fitted parameters.
\end{backmatter}


\bibliography{sample}

\noindent
\bigskip
\newpage

\section*{\centering Quantitative Decomposition of Speckle Decorrelation for Inverse Problems in Complex Wave Scattering: supplemental document}

\renewcommand\thefigure{S\arabic{figure}}    
\setcounter{figure}{0}  
\renewcommand{\theequation}{S\arabic{equation}}
\setcounter{equation}{0}

\tableofcontents

\section{Experimental methods and sample preparations}
    
\begin{figure}[htb!]
        \centering
        \includegraphics[width=0.6\linewidth]{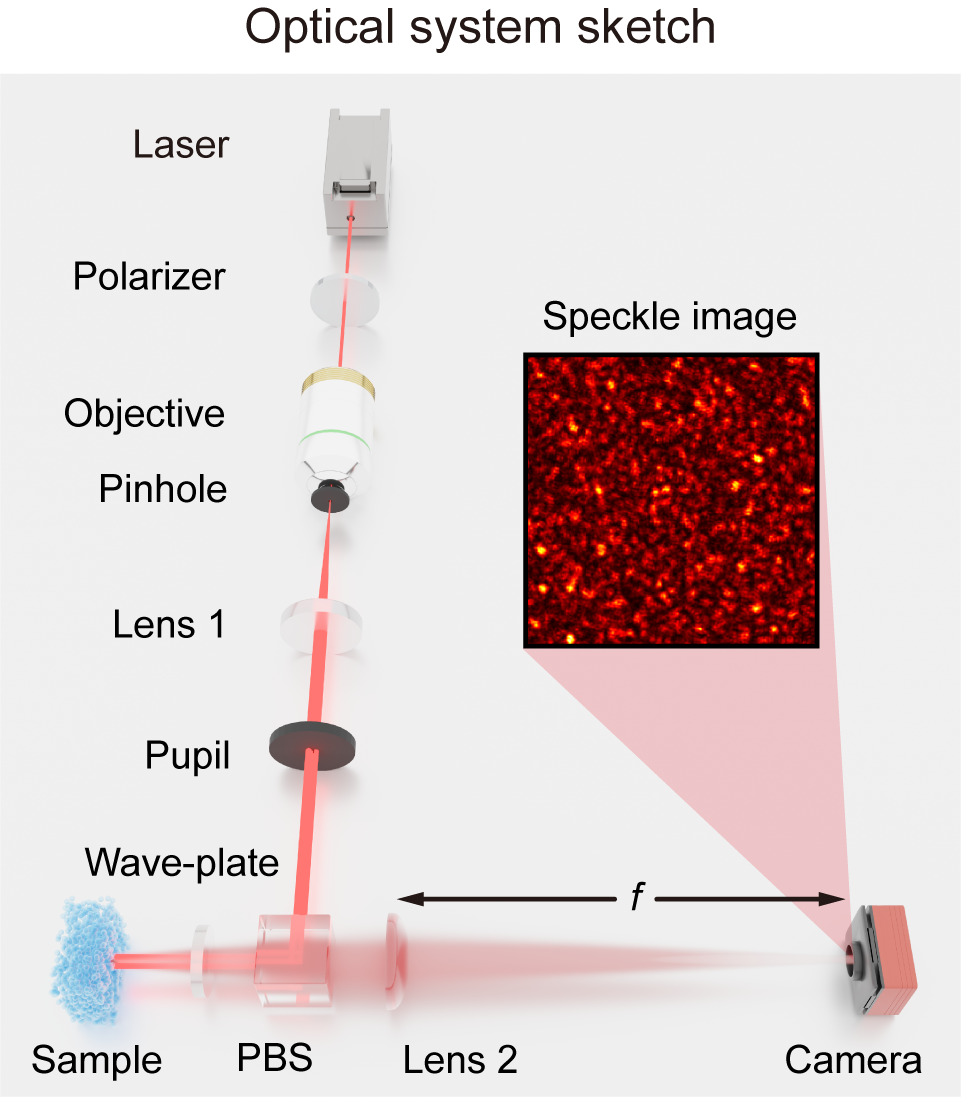}
        \caption{ A sketch of the optical scattering system.
        }
        \label{fig:s1}
    \end{figure}

    Fig.\ref{fig:s1} shows the optical system in this work to collect scattered light and demonstrate the decorrelation. A clean, collimated beam is produced after the Lens-1. A pupil reshapes the incident beam profile into a specific pattern, e.g. the Tai-Chi cat shape in Fig.1(a). The polarizing beam splitter (PBS) reflects the incident beam onto a potassium chloride (KCl) powder sample. After the polarization rotates 90 degrees by passing through the quarter wave-plate twice, the scattered beam transmits through the PBS and forms a speckle pattern at the focal plane of the Lens-2 with a focal length of $f=330$ mm. The inset shows an example of collected speckle patterns. 
    
    The laser model is Cobolt Flamenco 660 nm with 300 mW output power. The monochromatic camera model is QHY174, which contains 1920×1200 pixels with a 5.86 $\mu$m pixel size and 91 fps frame rate. A 800 $\mu$s exposure time was maintained to ensure a high quality image. The sample was placed on an translational stage with stepper motor (LTS150/M) to reduce the motion blur of speckle images. The effective distance from the sample to the pupil is 5 mm, close enough to ignore the diffraction of pupil patterns for a collimated beam. Commercially obtained potassium chloride (KCl) powder was used to calibrate the speckle. Samples of KCl with varying size distributions were prepared by sieving bulk KCl using sieves attached to the sieve shaker. The sieves were organized in decreasing order of opening size from top to bottom (Sieve opening sizes used: 500 $\mu$m, 450 $\mu$m, 355 $\mu$m, 300 $\mu$m, 250 $\mu$m, 200 $\mu$m, 106 $\mu$m). The sieving continued for 15-30 minutes until the powder weight in each sieves stabilized. Malvern Mastersizer 3000 attached to a Scirocco 2000 dry dispersion unit was used to obtain particle size data. The photon masks were fabricated by lithography, patterning 100 nm chromium on a 2.3 mm fused silica glass with a resolution of 0.25 $\mu$m.
    
    \section{Decorrelated memory effect in far field imaging systems}    
\begin{figure}[htb!]
        \centering
        \includegraphics[width=1\linewidth]{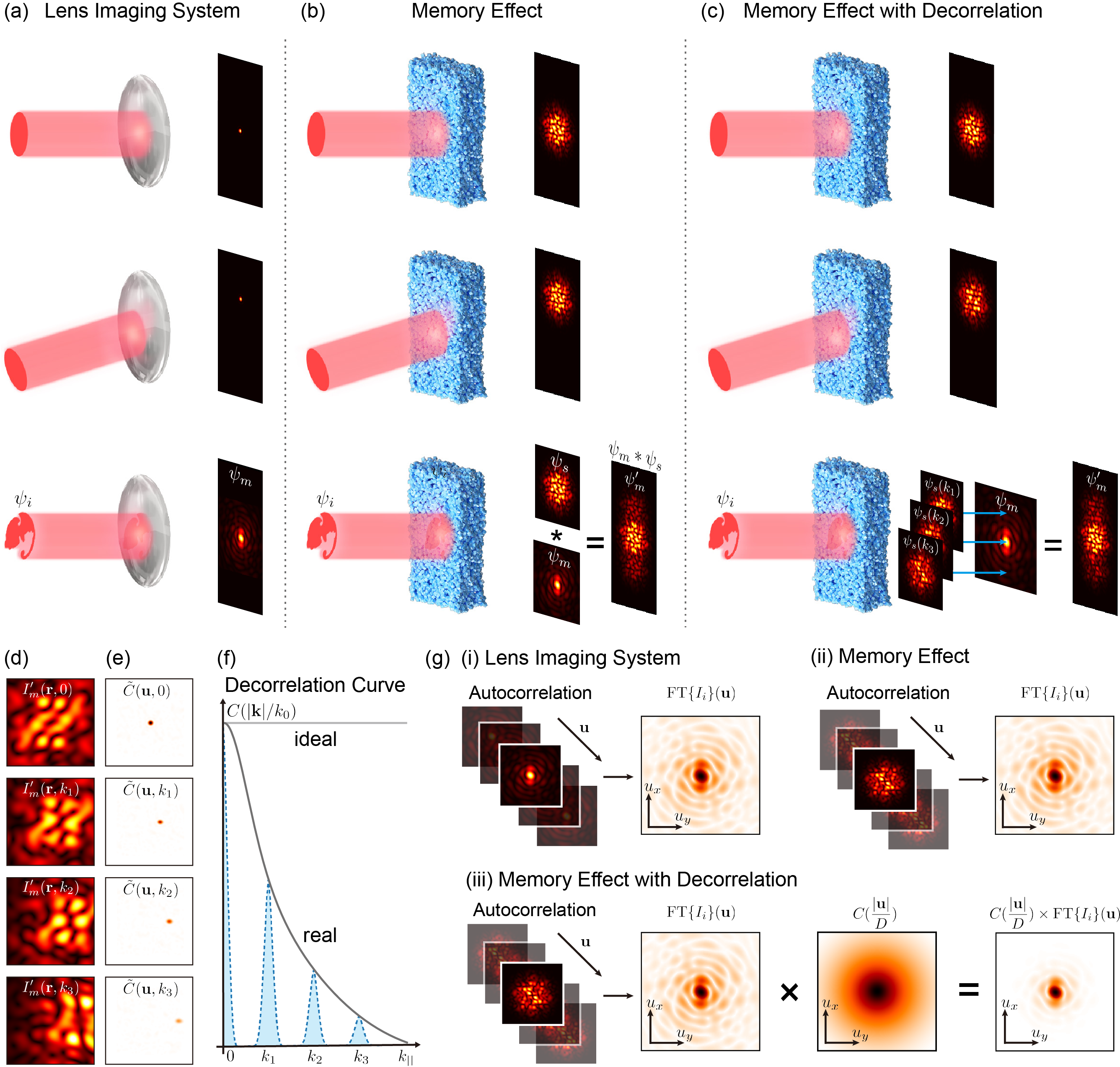}
        \caption{ Decorrelated memory effect in far field imaging systems. (a) Far-field imaging system with a single lens. The pattern detected at the focal plane is a convolution of incident field and the point-spread-function.
        (b) Far-field imaging system with a scattering sample. Idea memory effect guarantees the translational invariance of the system and defines the point-spread-function. 
        (c) Since the decorrelation breaks the translational symmetry in the system, the measured pattern is a summation of responses to different incident angles. 
        (d) Simulations of the scattered intensity from different incident directions. Apart from a translational shift, the pattern also deforms when $\mathbf{k}$ deviates from the original direction.
        (e) Cross-correlations between the speckle field from the straight illumination and the tilted illumination with an angle $|\mathbf{k}|/k_0$.
        (f) A sketch of the decorrelation curve. Blue dashed line represents cross-correlations for various $k$ values, their envelope outlines the normalized decorrelation curve (black solid line).
        (g) The field autocorrelation of the measured pattern is proportional to the power spectral density of the illumination profile for (i) lens imaging system and (ii) scattering imaging without decorrelation. (iii) The influence of the decorrelation is to introduce a low pass filter to the pupil’s power spectral density.
        }
        \label{fig:s2}
    \end{figure}

    Schematics of experiments for far-field coherent imaging, utilizing either a single lens or a scattering medium, as well as numerical examples, are presented in Fig.\ref{fig:s2}(a)-(c). The ideal single lens imaging system exhibits a translational invariance. It transforms a plane wave into a point, and a patterned incident field $\psi_i$ leads to its Fourier Transform $\psi_m=\text{FT}\{\psi_i\}$ at the far-field detector plane, as demonstrated in Fig.\ref{fig:s2}(a). According to the memory effect, imaging with a thin-layer scattering medium constitutes a shift-invariant system characterized by a speckle-shaped point-spread function $\psi_s\equiv \psi_s(r)$, as depicted in Fig.\ref{fig:s2}(b). The field at the detector plane $\psi'_m$ is a convolution of the Fourier Transform of the incident field and the speckle point-spread function, $\psi'_m =\text{FT}\{ \psi_i\} \ast \psi_s = \psi_m \ast \psi_s$, with a constant factor included for unit matching. Here the symbol $\ast$ represents a convolution operation. Polarization is disregarded for simplicity in this report. By applying the convolution theorem, we obtain the mutual intensity of $\psi'_m$,
    \begin{equation}\label{eq:ideal_ME_imaging_1}
    \begin{aligned}
    \left[\psi'_m \star \psi'_m \right] (\mathbf{u})=\left[(\psi_m \ast \psi_s)\star(\psi_m \ast \psi_s)\right](\mathbf{u})=\left[(\psi_m \star \psi_m)\ast(\psi_s \star \psi_s)\right](\mathbf{u}),
    \end{aligned}
    \end{equation} 
    where $\star$ denotes the spatial-integrated autocorrelation operation and $\mathbf{u}$ is the displacement coordinate in the autocorrelation. Because the speckle field at the detector plane is essentially a random Gaussian variable, the mutual intensity of the speckle field, denoted as $\psi_s \star \psi_s$, approximates a sharp peak\cite{Katz2014}, akin to a $\delta$-function. Consequently, the right-hand side of equation.(\ref{eq:ideal_ME_imaging_1}) reduces to $\psi_m \star \psi_m$. According to the Gaussian momentum theorem\cite{Goodman2015}, a straightforward relationship exists between the intensity autocorrelation and the field mutual intensity, which is $I \star I =\overline{I}^2 + |\psi \star \psi|^2$, where $\overline{I}^2$ represents the mean intensity over the detector. Hence, as shown in Fig.\ref{fig:s2}(g)(i) and (ii), within the ideal memory effect range, the autocorrelation of the measured intensity $I'_m$ is equivalent to the power spectral intensity of the incident beam intensity $I_i$, apart from a constant factor,
    
    \begin{equation}\label{eq:ideal_ME_imaging_2}
    \begin{aligned}
    I'_m \star I'_m - \overline{I'_m}^2 &= |\psi'_m \star \psi'_m|^2 =|\psi_m \star \psi_m|^2 \\
    &= |\text{FT}\{ \psi_i\} \star \text{FT}\{ \psi_i\}|^2 = |\text{FT}\{ |\psi_i|^2 \}|^2=|\text{FT}\{ I_i \}|^2.
    \end{aligned}
    \end{equation}

    However, in many practical scenarios, the memory effect decorrelates with an increased tilt, resulting in a restricted effective angle, as illustrated by the simulated images in Fig.\ref{fig:s2}(c) and Fig.\ref{fig:s2}(d). Here, $\psi_s\equiv \psi_s(\mathbf{r},\mathbf{k})$ is no longer shift-invariant and acquires an angle dependence, with $\mathbf{k}$ representing the in-plane component of the wave vector. In Fig.\ref{fig:s2}(e), $\tilde C(\mathbf{u},\mathbf{k})=[\psi_s(\mathbf{r},\mathbf{0}) \star \psi_s(\mathbf{r},\mathbf{k}) ](\mathbf{u})$ is the correlation map between the speckle field from the straight illumination and the tilted illumination with an angle $\theta=|\mathbf{k}|/k_0$. $k_0$ is the wave vector of the incident beam. Blue dashed line in Fig.\ref{fig:s2}(f) represents  cross-correlations for various $\mathbf{k}$ values and their envelope outlines the normalized decorrelation curve $C(\theta=|\mathbf{k}|/k_0)=
    \frac{\tilde C(\frac{D}{k_0}\mathbf{k},\mathbf{k})}{\tilde C(\textbf{0},\textbf{0})}$, where $D$ is the far-field distance between the scattering medium and the detector. Therefore, the electric field $\psi'_m$ at the detector plane is a linear combination of $\psi_s(\mathbf{r},\textbf{k})$.
    \begin{equation}\label{eq:real_ME_1}
    \begin{aligned}
    \psi'_m(\mathbf{r}) = \int \psi_m \left(\frac{D}{k_0}\mathbf{k}\right)\psi_s(\mathbf{r},\mathbf{k})\text{d}^2\mathbf{k}.
    \end{aligned}
    \end{equation} 
    
    Upon taking its mutual intensity, we obtain:       \begin{equation}\label{eq:real_ME_2}
    \begin{aligned}
    &\left[\psi'_m \star \psi'_m\right](\mathbf{u})\\
    &= \int \psi_m(\frac{D}{k_0}\mathbf{k}) \psi_s(\mathbf{r},\mathbf{k}) \psi_m^{\ast}(\frac{D}{k_0}(\mathbf{k}+\Delta \mathbf{k}))\psi_s^{\ast}(\mathbf{r+u},\mathbf{k}+\Delta \mathbf{k} )\text{d}^2\mathbf{k}\text{d}^2\Delta \mathbf{k}\text{d}^2\mathbf{r}\\
    &= \int \text{d}^2\Delta k \left( \int \psi_m(\frac{D}{k_0}\mathbf{k}) \psi_m^{\ast}(\frac{D}{k_0} (\mathbf{k}+\Delta \mathbf{k}) ) \text{d}^2\mathbf{k}\right) \left( \int \psi_s(\mathbf{r},\mathbf{k}) \psi_s^{\ast}(\mathbf{r+u},\mathbf{k}+\Delta \mathbf{k} )\text{d}^2\mathbf{r} \right) \\  
    &=\int \left[\psi_m \star \psi_m\right](\frac{D}{k_0}\Delta \mathbf{k}) \tilde C(\mathbf{u},\Delta \mathbf{k})\text{d}^2\Delta \mathbf{k}\\
    &=\int \left[\psi_m \star \psi_m\right](\frac{D}{k_0}\Delta \mathbf{k})  C\left(\frac{|\Delta \mathbf{k}|}{k_0} \right) \delta(\mathbf{u}-\frac{D}{k_0} \Delta \mathbf{k})\text{d}^2\Delta \mathbf{k}\\
    &= \left[\psi_m \star \psi_m\right](\mathbf{u})  C\left(\frac{\mathbf{u}}{D} \right).
    \end{aligned}
    \end{equation} 
    
    Thus, as shown in Fig.\ref{fig:s2}(g)(iii), the autocorrelation of the measured intensity $I'_m$ is a multiplication of the decorrelation curve $C$ and the power spectral density of the incident beam $\left|\text{FT}\{ I_i \}(\mathbf{u})\right|^2$, with an additional constant $\overline{I'_m}^2$ originated from the spatially averaged intensity at the detector plane, 
    \begin{equation}\label{eq:real_ME_3}
    \begin{aligned}
    \left[ I'_m \star I'_m\right](\mathbf{u}) -\overline{I'_m}^2 & = | \psi'_m \star \psi'_m  |^2  (\mathbf{u})=\left|\text{FT}\{ I_i \}(\mathbf{u})\right|^2  \left|C\left(\frac{\mathbf{u}}{D} \right)\right|^2.
    \end{aligned}
    \end{equation} 

    This expression shows the influence of the decorrelation on the coherent far-field imaging system, essentially introducing a low pass filter to the pupil's power spectral density.
    
    \section{Decomposition principle of the ensemble autocorrelation} 
    Speckle autocorrelations encode information from both incident beams and scattering media. The first term in equation.(\ref{eq:real_ME_3}), $\left|\text{FT}\{ I_i \}(\mathbf{u})\right|^2$, is fully determined by the incident beam. Therefore, without conducting any mathematical operations,  it is evident that details about the scattering medium are embedded in the remaining term, specifically, the decorrelation curve $\left| C\left(\frac{\mathbf{u}}{D} \right)\right|^2$. In this section, we will proof the main conclusion of the decomposition principle: the decorrelation in scattered light is a superposition of contributions from thin-layer and volumetric scattering, providing explicit expressions for terms in equation.(1).

    \subsection{Decomposition of speckle decorrelations}

    Assessing $C$ in the reflection mode is a difficult task, particularly for thick materials or large particles whose size greatly exceeds the optical wavelength. This complexity arises because the multi-scattering effect surpasses the boundaries of the first-Born approximation. 
    To manage the multiple scattering challenge at each particle, we turned to the first principle Lippmann-Schwinger equation (LSE), derived from Maxwell's equations,    
    \begin{equation}\label{eq:LippSch}
    \begin{aligned}
    \psi(\mathbf{r})& = \psi_{i}(\mathbf{r})+ \psi_{s}(\mathbf{r}) \\
    &=\psi_{i}(\mathbf{r})+ k_0^2
    \int G(\mathbf{r}-\mathbf{r'})V(\mathbf{r'})
    \psi(\mathbf{r'})\:\text{d}^3\mathbf{r'}.
    \end{aligned}
    \end{equation}  
    This integral formula establishes a relationship between the total field $\psi$, the incident field $\psi_{i}$, the scattering potential $V(\mathbf{r}) \equiv  n(\mathbf{r})^2-1$, and the free-space Green's function $G=\frac{e^{ik_0r}}{4\pi r}$. However, obtaining its numerical solution demands extensive computational power for macroscopic scattering systems, given that the scalar field $\psi$ appears on both side of the equation\cite{Pang2022}. Instead, we are primarily interested in its second order statistical properties, namely, the mutual intensity and the autocorrelation.
    
    \begin{figure}[htb]
    \centering
    \includegraphics[width=0.6\linewidth]{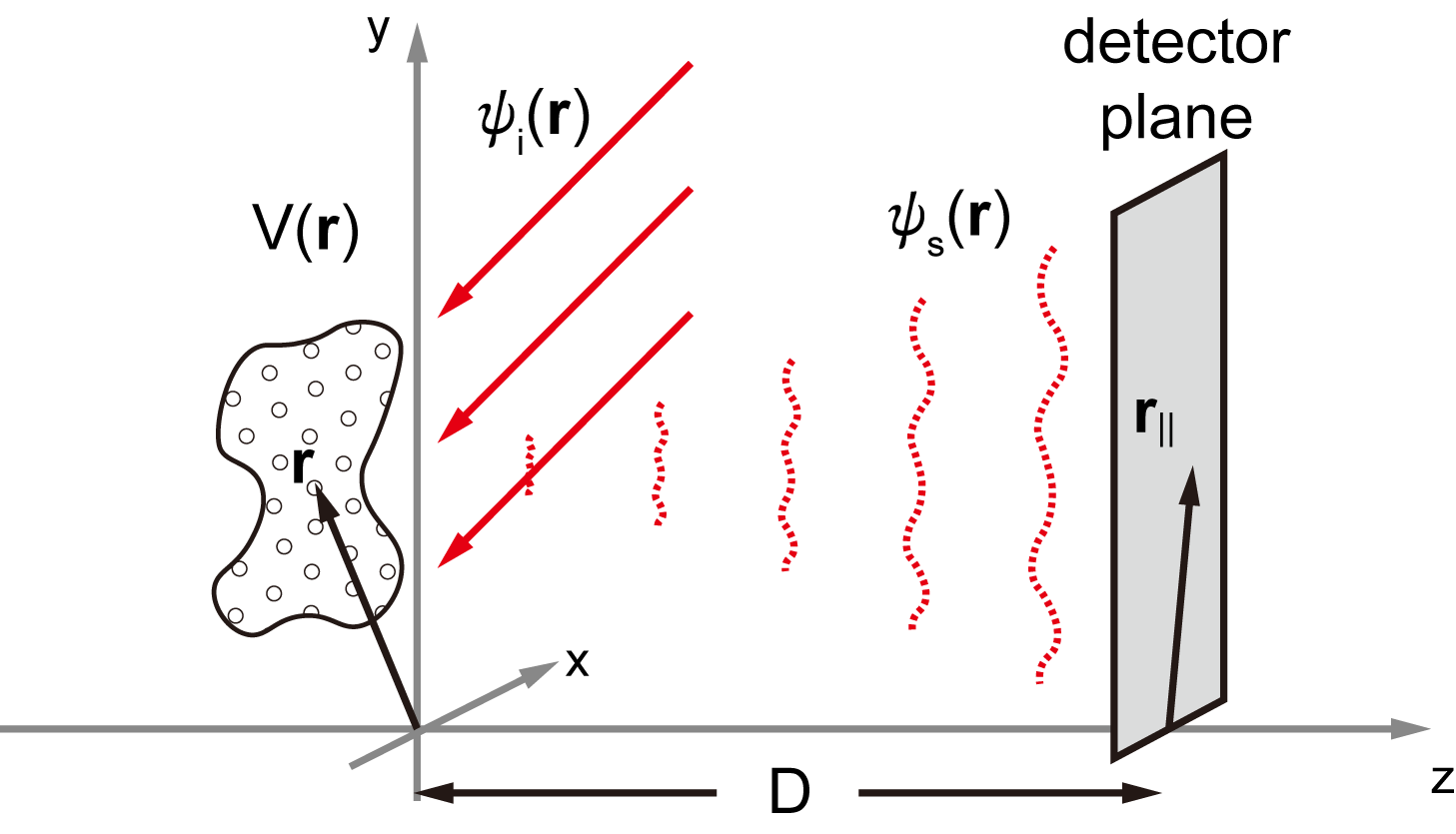}
    \caption{ Sketch of the speckle measurement system. Sketch of the speckle measurement system. We assume that only back-scattered light is collected by the detector and there is no reflective substrate at left of scatterers }
    \label{fig:s3}
    \end{figure}
    
    As shown in Fig.\ref{fig:s3}, Since the incident beam has no contribution at the far-field plane where the detector is located, the total field and its mutual intensity at \textbf{the detector plane} are 
    \begin{equation}\label{eq:farfield-E}
    \begin{aligned}
    \psi(\mathbf{r})& = \frac{k_0^2}{D}e^{ik_0D+ik_0\frac{r_{||}^2}{2D}}\int e^{-\frac{ik_0}{D}\mathbf{r}_{||}\cdot \mathbf{r}'_{1||}} e^{\frac{ik_0}{2D}{r^{\prime 2}_{1}}+ik_0r'_{1z}}V(\mathbf{r}'_1)\psi(\mathbf{r}'_1)\text{d}^3\mathbf{r}'_1
    \end{aligned}
    \end{equation}  
    
    \begin{equation}\label{eq:farfield-Edouble}
    \begin{aligned}
    \psi(\mathbf{r}) \psi^{\ast}(\mathbf{r+u})& = \frac{k_0^4}{D^2}e^{\frac{ik_0}{2D}\left[ r_{||}^2-(r_{||}+u_{||})^2 \right]} \\
    &\int e^{\frac{ik_0}{D}\left[ \mathbf{r}_{||}\cdot (\mathbf{r}'_{1||}-\mathbf{r}'_{2||}) - \mathbf{u}_{||}\cdot \mathbf{r}'_{2||}\right]} e^{\frac{ik_0}{2D}({r^{\prime 2}_{1}}-{r^{\prime 2}_{2}})} e^{ik(r'_{1z}-r'_{2z})}\\ &V(\mathbf{r}'_1)\psi(\mathbf{r}'_1)V(\mathbf{r}'_2)\psi^{\ast}(\mathbf{r}'_2)\text{d}^2\mathbf{r}'_{1||} \text{d}r'_{1z} \text{d}^2\mathbf{r}'_{2||} \text{d}r'_{2z},
    \end{aligned}
    \end{equation} 
    where the integral over $\mathbf{r'}$ is restricted to the scattering volume where $V(\mathbf{r})\neq0$.
    
    As our detector is typically a 2D array, variables $\mathbf{r}$ and $\mathbf{u}$ only have parallel components, so we ignore the parallel subscript in the rest text. If we use a lens with a focal length $f=D$ to mimic the far-field case at the distance $D$, the same result should appear, except that the $e^{\frac{ik_0}{2f}\left[ r_{||}^2-(r_{||}+u_{||})^2 \right]}$ term will be omitted and the full result is independent to $r_{||}$. Thus, we substitute $D$ with $f$ in the following text. The spatial-integral mutual intensity $\mu=\psi \star \psi$ is,
    \begin{equation}\label{eq:farfield-cor}
    \begin{aligned}
    \mu(\mathbf{u})\equiv\int \psi(\mathbf{r}) \psi^{\ast}(\mathbf{r+u}) \text{d}^2\mathbf{r}
    & = \frac{k_0^4}{f^2}
    \int \delta (\mathbf{r}'_{1||}-\mathbf{r}'_{2||}) e^{\frac{ik_0}{f} \mathbf{u}\cdot \mathbf{r}'_{2||}} e^{\frac{ik_0}{2f}({r^{\prime 2}_{1}}-{r^{\prime 2}_{2}})} \\ 
    &e^{ik_0(r'_{1z}-r'_{2z})}V(\mathbf{r}'_1)\psi(\mathbf{r}'_1)V(\mathbf{r}'_2)\psi^{\ast}(\mathbf{r}'_2)\text{d}^2\mathbf{r}'_{1||} \text{d}r'_{1z} \text{d}^2\mathbf{r}'_{2||} \text{d}r'_{2z}\\
    &= \frac{k_0^4}{f^2}
    \int  e^{\frac{ik_0}{f} \mathbf{u}\cdot \mathbf{r}'_{1||}} e^{\frac{ik_0}{2f}({r^{\prime 2}_{1z}}-{r^{\prime 2}_{2z}})} e^{ik_0(r'_{1z}-r'_{2z})}\\ &V\psi(\mathbf{r}'_{1||},r'_{1z})V\psi^{\ast}(\mathbf{r}'_{1||},r'_{2z})\text{d}^2\mathbf{r}'_{1||} \text{d}r'_{1z} \text{d}r'_{2z},
    \end{aligned}
    \end{equation}   
    where $V\psi(\cdot) \equiv V(\cdot)\psi(\cdot)$. Furthermore, after defining the demodulation notation $V\tilde \psi (\mathbf{r}'_{1||},r'_{1z}) \equiv V(\mathbf{r}'_{1||},r'_{1z})\psi(\mathbf{r}'_{1||},r'_{1z})e^{ik_0r'_{1z}}$, we obtain,
    \begin{equation}\label{eq:mutual_intensity}
    \begin{aligned}
    \mu(\mathbf{u}
    ) & = \frac{k_0^4}{f^2}
    \int  e^{\frac{ik_0}{f} \mathbf{u}\cdot \mathbf{r}'_{1||}} \\ 
    & \left(\int e^{\frac{ik_0}{2f}({r^{\prime 2}_{1z}}-{r^{\prime 2}_{2z}})} V\tilde\psi(\mathbf{r}'_{1||},r'_{1z})V\tilde\psi^{\ast}(\mathbf{r}'_{1||},r'_{2z})   \text{d}r'_{1z} \text{d}r'_{2z} \right) \text{d}^2\mathbf{r}'_{1||}.
    \end{aligned}
    \end{equation} 
    
    Therefore, $\mu(\mathbf{u})$ is the 2D-Fourier transform of the term in the bracket, which we denote as $s(\mathbf{r}'_{1||})$.
    \begin{equation}\label{eq:u_s}
    \begin{aligned}
    \mu(\mathbf{u}
    ) = \text{FT}_{\text{2D}}\{ s(\mathbf{r}'_{1||})\}(\mathbf{u}
    ) 
    \end{aligned}
    \end{equation} 
    \begin{equation}\label{eq:s_total}
    \begin{aligned}
    s(\mathbf{r}'_{1||}) & = \int e^{\frac{ik_0}{2f}({r^{\prime 2}_{1z}}-{r^{\prime 2}_{2z}})} V\tilde\psi(\mathbf{r}'_{1||},r'_{1z})V\tilde\psi^{\ast}(\mathbf{r}'_{1||},r'_{2z})  \text{d}r'_{1z} \text{d}r'_{2z}.
    \end{aligned}
    \end{equation}

    The mutual intensity in a single realization can be rewritten as,
    \begin{equation}\label{eq:single_mu}
    \begin{aligned}
    \mu(\mathbf{u})=\left< \mu(\mathbf{u}) \right>+\delta\mu(\mathbf{u}),
    \end{aligned}
    \end{equation} 
    where $\left< \mu \right>$ is in the ensemble sense and $\delta \mu$ indicates the zero-mean fluctuation induced by the single realization. Due to the circular complex Gaussian statistics of the field, we obtain a simple relation,
    \begin{equation}\label{eq:single_auto}
    \begin{aligned}
    \left[ I \star I\right](\mathbf{u}) &=\left \langle I \right \rangle ^2  + | \mu (\mathbf{u})|^2\\
    &=\left \langle I \right \rangle ^2  + | \left<\mu(\mathbf{u})\right> |^2+ |\delta \mu(\mathbf{u})|^2 + 2\text{Re}\{ \mu(\mathbf{u}) \delta \mu^{\ast}(\mathbf{u})\}\\
    \end{aligned}
    \end{equation}
    where $\left[ I \star I\right](\mathbf{u})$ is defined as the spatial-integral intensity autocorrelation. $\left \langle I \right \rangle ^2$ is the mean intensity, which is a constant and independent to $\mathbf{u}$. Upon taking the ensemble of $\left[ I \star I\right](\mathbf{u})$, we obtain    
    \begin{equation}\label{eq:ensemble_auto}
    \begin{aligned}
    \left< \left[ I \star I\right](\mathbf{u}) \right> &\equiv \left \langle \int I( \mathbf{r})I^{\ast}( \mathbf{r+u})\:\text{d}^2\mathbf{r} \right \rangle =\left \langle I \right \rangle ^2  + |\left< \mu(\mathbf{u}) \right>|^2+\left< |\delta\mu( \mathbf{u})| ^2 \right>,
    \end{aligned}
    \end{equation}
    Since $|\delta \mu|$ is much smaller than the averaged $\left< \mu \right>$, we can safely ignore the former one. Also, according to the ergodicity in most scattering systems, we can substitute 
    the ensemble average $\left \langle \cdot \right\rangle$ with the temporal average $\left \langle \cdot \right\rangle_t$ in the experiment. Thus, our observation $\left< \left[ I \star I\right](\mathbf{u}) \right>=\left \langle I \right \rangle ^2 + |\left< \mu(\mathbf{u}) \right>|^2$. In this section, we focus on the ensemble $\left< \mu \right>$, and we will return to $\delta \mu$ in the last section. Considering the equation.(\ref{eq:u_s}) and equation.(\ref{eq:s_total}) in the ensemble sense, 
    \begin{equation}\label{eq:ensemble_u}
    \begin{aligned}
    \left< \mu(\mathbf{u})\right>= \text{FT}_{\text{2D}}\{ \langle s(\mathbf{r}'_{1||}) \rangle \}(\mathbf{u}
    ) 
    \end{aligned}
    \end{equation} 
    \begin{equation}\label{eq:ensemble_s}
    \begin{aligned}
    \langle s(\mathbf{r}'_{1||})\rangle & = \int e^{\frac{ik_0}{2f}({r^{\prime 2}_{1z}}-{r^{\prime 2}_{2z}})} \left \langle V\tilde\psi(\mathbf{r}'_{1||},r'_{1z})V\tilde\psi^{\ast}(\mathbf{r}'_{1||},r'_{2z}) \right \rangle_t \text{d}r'_{1z} \text{d}r'_{2z}.
    \end{aligned}
    \end{equation}

    Here two qualitatively distinct scenarios are necessary to be discussed, $r'_{1z}\neq r'_{2z}$ (thin-layer scattering) and $r'_{1z}\approx r'_{2z}$(volumetric scattering). 

    \textbf{1. Thin-layer scattering,} $\mathbf{r'_{1z}\neq r'_{2z}}$
    
    When $r'_{1z}\neq r'_{2z}$, the integral over $r'_{1z}$ and $r'_{2z}$ are independent. Moreover, $e^{-\frac{ik_0}{2f}{r^{\prime 2}_{z}}}$ has strong oscillation at high $r'_{z}$ region, while the rest term is relatively smooth. So the integral over $r'_z$ is approximately confined within a thin-layer range $\Delta = \sqrt{\pi f/k_0}$. In our optical system, $f\approx 33$ cm and $\lambda=660$ nm, so $\Delta$ is approximately 330 $\mu$m, which corresponds to 1-3 layers particles in our interested particle size range 100 $\mu$m - 500 $\mu$m. Thus, we can safely apply the first Born approximation when $r'_{1z}\neq r'_{2z}$, the thin-layer contribution $\langle s_{\text{2D}} \rangle$ is, 
        
    \begin{equation}\label{eq:s_2D_inter}
    \begin{aligned}
    \langle s_{\text{2D}}(\mathbf{r}'_{1||}) \rangle & =  \left \langle \int V\tilde\psi(\mathbf{r}'_{1||},r'_{1z})e^{\frac{ik}{2f}r^{\prime 2}_{1z}}\text{d}r'_{1z} \int V\tilde\psi^{\ast}(\mathbf{r}'_{1||},r'_{2z})e^{-\frac{ik}{2f}r^{\prime 2}_{2z}} \text{d}r'_{2z}\right \rangle_t \\
    &\approx  \left \langle \tilde\psi_i(\mathbf{r}'_{1||})\tilde\psi_i^{\ast}(\mathbf{r}'_{1||})\int V(\mathbf{r}'_{1||},r'_{1z})e^{\frac{ik}{2f}r^{\prime 2}_{1z}}\text{d}r'_{1z} \int V(\mathbf{r}'_{1||},r'_{2z})e^{-\frac{ik}{2f}r^{\prime 2}_{2z}} \text{d}r'_{2z}\right \rangle_t \\
    &\approx  \left \langle \frac{\pi f}{k} I_i(\mathbf{r}'_{1||}) |\langle V (\mathbf{r}'_{1||})\rangle_z |^2  \right \rangle _t\\
    &=  \frac{\pi f}{k} I_i(\mathbf{r}'_{1||})\left \langle |\langle V (\mathbf{r}'_{1||})\rangle_z |^2  \right \rangle _t.
    \end{aligned}
    \end{equation} 
    
    Here we want to particularly point out that $\langle \cdot \rangle_z$ is the z-direction spatial average over the top thin layer with a thickness $\Delta$. $\psi_i$ and $I_i$ are the incident field and intensity distribution. 
    
    \textbf{2. Volumetric scattering, }$\mathbf{r'_{1z}\approx r'_{2z}}$
    
    When $r'_{1z} \approx r'_{2z}$, more precisely, $|r'_{1z}-r'_{2z}|<\epsilon(r'_z)$, no matter at which $r'_{1z}$ value, $e^{\frac{ik_0}{2f}(r^{\prime 2}_{z1}-r^{\prime 2}_{z2})}$ term contributes unity, the integral over $r'_{1z}$ can spread over the whole scattering volume. The value $\epsilon(r'_z)$ will be further discussed later. This is corresponding to the volumetric contribution, 
    
    \begin{equation}\label{eq:s_2_1}
    \begin{aligned}
    \langle s_{\text{3D}}(\mathbf{r}'_{1||}) \rangle & =  \int\epsilon(r'_{1z}) \left \langle V\tilde\psi(\mathbf{r}'_{1||},r'_{1z})V\tilde\psi^{\ast}(\mathbf{r}'_{1||},r'_{1z}) \right \rangle_t \text{d}r'_{1z} \\
    & =  \int \epsilon(r'_{1z}) \left \langle V^2(\mathbf{r}'_{1||},r'_{1z})I(\mathbf{r}'_{1||},r'_{1z}) \right \rangle_t \text{d}r'_{1z}.
    \end{aligned}
    \end{equation}

    At large $r'_z$, electrical field is strongly scattered, we assume $I(\mathbf{r}'_{1||},r'_{1z})$ varies much faster than $V^2(\mathbf{r}'_{1||},r'_{1z})$. Moreover, as later discussed in the diffusion theory, $\langle I \rangle_t$ varies much slower than $V$. Therefore, we obtain the following formula by applying the rotating-wave approximation twice,
    
    \begin{equation}\label{eq:s_2_2}
    \begin{aligned}
    \langle s_{\text{3D}}(\mathbf{r}'_{1||}) \rangle & =  \int \epsilon(r'_{1z}) \left \langle V^2(\mathbf{r}'_{1||},r'_{1z})I(\mathbf{r}'_{1||},r'_{1z}) \right \rangle_t \text{d}r'_{1z}\\
    & = \int \epsilon(r'_{1z}) \left \langle V^2(\mathbf{r}'_{1||},r'_{1z})  \langle I(\mathbf{r}'_{1||},r'_{1z}) \rangle_t \right \rangle_t
    \text{d}r'_{1z}\\
    &=  \int \epsilon(r'_{1z}) \left \langle V^2(\mathbf{r}'_{1||},r'_{1z})\right \rangle_t \left \langle I(\mathbf{r}'_{1||},r'_{1z}) \right \rangle_t \text{d}r'_{1z}.
    \end{aligned}
    \end{equation}
    $\langle V^2(\mathbf{r}'_{1||},r'_{1z})\rangle_t=\gamma(n^2-1)^2$ is a constant, where $\gamma$ is the volume filling-factor of scatters. $ \langle I(\mathbf{r}'_{1||},r'_{1z}) \rangle_t$ is well described by diffusion theory, which will be discussed later. Thus, 
    \begin{equation}
    \begin{aligned}
    \langle s_{\text{3D}}(\mathbf{r}'_{1||}) \rangle
    &=  \gamma(n^2-1)^2\int \epsilon(r'_{1z})  \left \langle I(\mathbf{r}'_{1||},r'_{1z}) \right \rangle_t \text{d}r'_{1z}.
    \end{aligned}
    \end{equation}
    
    Additionally, we discuss the value of $\epsilon$. As the wavefront is strongly distorted in the multi-scattering region, $ \langle V\tilde\psi(\mathbf{r}'_{1||},r'_{1z})V\tilde\psi^{\ast}(\mathbf{r}'_{1||},r'_{2z})  \rangle_t$ in equation.(\ref{eq:s_2_2}) will quickly vanish to zero if $|r'_{1z}-r'_{2z}|>\epsilon(r'_{1z})$, where $\epsilon(r'_{1z})$ is denoted as the phase decorrelation depth along the z-direction. Without a rigid proof, we simply assume that this decorrelation depth is independent to particle size and sample thickness. The rationality of this assumption is validated by the fitting result in the next section. Therefore, $\epsilon$ could be moved out of the integral symbol,
    
    \begin{equation}\label{eq:s_3D_inter}
    \begin{aligned}
    \langle s_{\text{3D}}(\mathbf{r}'_{1||}) \rangle
    &=  (n^2-1)^2 \gamma \epsilon  \int  \left \langle I(\mathbf{r}'_{1||},r'_{1z}) \right \rangle_t \text{d}r'_{1z}.
    \end{aligned}
    \end{equation}
    
    The integration over the z-direction indicates the contribution of $\langle s_{\text{3D}} \rangle$ comes from the decorrelation of volumetric scattering deeply into the scattering medium. 
    
    \textbf{3. Decomposition principle}

    Based on the above discussions, we proved the decomposition principle for the mutual intensity $\left[\psi \star \psi\right](\mathbf{u})$,
    \begin{equation} \label{eq:decom_principle}
    \begin{aligned}
    \langle \mu(\mathbf{u}) \rangle& = \text{FT}_{\text{2D}} \{ \langle s_{\text{2D}}(\mathbf{r}'_{1||})\rangle + \langle s_{\text{3D}}(\mathbf{r}'_{1||}) \rangle \}(\mathbf{u}).\\
    \end{aligned}
    \end{equation} 
    Explicit expressions of contributions from thin-layer scattering and volumetric scattering are discussed in the following sections, respectively.
    
    \subsection{Thin-layer decorrelation}
    
    According to the discussion about thin-layer scattering in our previous work\cite{Zhang2023}, the contribution from $\langle s_{\text{2D}} \rangle$ is, 

    \begin{equation}\label{eq:FTs1}
    \begin{aligned}
    \text{FT}_{\text{2D}}\{ \langle s_{\text{2D}}(\mathbf{r}'_{1||}) \rangle\}(\mathbf{u}
    ) &= (n^2-1)^2 \frac{\pi f}{k_0} \text{FT}_{\text{2D}}\{ I_i \} \text{FT}_{\text{2D}}\{  \langle a({\rho})\rangle_{\rho}\}\\
    &=(n^2-1)^2 \frac{\pi f^2}{k_0^2} \text{FT}_{\text{2D}}\{ I_i \} \frac{\int \pi \rho^2 \text{Jinc}(\frac{\rho k_0 |\mathbf{u}|}{f})p(\rho)\text{d}\rho}{\int \pi \rho^2 p(\rho)\text{d}\rho},
    \end{aligned}
    \end{equation} 
    where $I_i(\mathbf{r}'_{||})$ is the incident beam profile, $n$ is the refractive index of scatters. $a({\rho})$ is the scatter shape as a function of particle size ${\rho}$, $a({\rho})=1$ if the particle size is larger than $\rho$, otherwise $a({\rho})=0$. $\text{Jinc}(x)=\frac{2J_1(x)}{x}$ is the Sombrero function, where $J_1(x)$ is the Bessel function of the first kind and first order. $p(\rho)$ is the probability distribution of the particle size $\rho$.

    \subsection{Volumetric decorrelation}
    According to equation.(\ref{eq:s_3D_inter}), interference features of the intensity distribution $\langle I(\mathbf{r}') \rangle_t$ are eliminated by the ensemble average, so that the diffusion theory is enough to estimate its behaviors. Considering a random walk model, we set the step size $\rho_z$ to be the mean-free-path, which is approximately the mean particle size for the dense powder,
    \begin{equation}
    \begin{aligned}
    \rho_z = \int \rho p(\rho)\text{d} \rho.
    \end{aligned}
    \end{equation}
    The corresponding lateral step size $\rho_l$ is,
    \begin{equation}
    \begin{aligned}
    \rho_l = \langle \tan{\Delta \theta} \rangle \rho_z = \langle \tan{\Delta \theta} \rangle \int \rho p(\rho)\text{d} \rho,
    \end{aligned}
    \end{equation}
    where $\Delta \theta$ is the deflection angle for a single scattering event. $\langle \tan{\Delta \theta} \rangle$ is related to the anisotropy parameter $g$, the average cosine of the deflection angle.  

    After $N$ steps from the incident point, the penetration depth is $z=N\rho_z$, and the non-dissipative intensity distribution $h(\mathbf{r_{||}},z)$ is,
    \begin{equation}
    \begin{aligned}
    h(\mathbf{r_{||}},z) =\frac{1}{2\pi N \rho_l^2}e^{-\frac{r^2_{||}}{2N \rho_l^2}}= \frac{1}{2\pi \rho_l z \langle \tan{\Delta \theta} \rangle}e^{-\frac{r^2_{||}}{2\langle \tan{\Delta \theta} \rangle \rho_l z}}
    \end{aligned}
    \end{equation}
    In practive, due to the back-scattering, a small dissipation ratio $\beta$ at each scattering event is inevitable, even if the material has no absorption at the incident wavelength. The dissipative intensity point-spread-function $h(\mathbf{r_{||}},z)$ and its 2D Fourier transform are, 
    \begin{equation}\label{eq:h_psf}
    \begin{aligned}
    h(\mathbf{r_{||}},z) =\frac{1}{2\pi N \rho_l^2}e^{-\frac{r^2_{||}}{2N\rho_l^2}} e^{-\beta N}= \frac{1}{2\pi \rho_l z\langle \tan{\Delta \theta} \rangle}e^{-\frac{r^2_{||}}{2\langle \tan{\Delta \theta} \rangle \rho_l z }}e^{- \frac{\beta z}{\rho_z}}.
    \end{aligned}
    \end{equation}
    \begin{equation}
    \begin{aligned}
    \text{FT}_{\text{2D}}\{h(\mathbf{r_{||}},z) \}(\mathbf{u})&= \int h(\mathbf{r_{||}},z) e^{i\frac{k_0}{f}\mathbf{u}\cdot \mathbf{r}_{||}} \text{d}^2\mathbf{r}_{||}\\
    &=  \frac{f}{k}e^{-( \frac{\beta}{\rho_z}+ \frac{\langle \tan{\Delta \theta} \rangle \rho_l k_0^2 u^2}{2f^2}) z}
    \end{aligned}
    \end{equation}
    The 3D ensemble-averaged intensity distribution is the 2D convolution of the incident beam profile $I_i(\mathbf{r}'_{||})$ and the intensity point spread function,
    \begin{equation}\label{eq:conv_of_h}
    \begin{aligned}
    \langle I(\mathbf{r'_{||}},z) \rangle_t = I_i(\mathbf{r}'_{||}) \ast h(\mathbf{r}'_{||},z) 
    \end{aligned}
    \end{equation}
    Thus, bringing equation.(\ref{eq:h_psf}) and equation.(\ref{eq:conv_of_h}) in equation.(\ref{eq:s_3D_inter}), we obtain the decorrelation from the volumetric scattering, 
    \begin{equation}
    \begin{aligned}
    \langle s_{\text{3D}}(\mathbf{r}'_{1||}) \rangle
    &=  \gamma(n^2-1)^2 \epsilon \int  \left \langle I_i(\mathbf{r}'_{||}) \ast h(\mathbf{r}'_{||},r'_z) \right \rangle_t \text{d}r'_{z},
    \end{aligned}
    \end{equation}
    \begin{equation}\label{eq:FTs2_inter}
    \begin{aligned}
    \text{FT}_{\text{2D}}\{ \langle s_{\text{3D}}(\mathbf{r}'_{1||}) \rangle\}(\mathbf{u}
    ) &= \gamma(n^2-1)^2 \epsilon \text{FT}_{\text{2D}}\{I_i\} \int \text{FT}_{\text{2D}}\{ h \}(\mathbf{u},r'_z)\text{d}r'_{z}.\\
    \end{aligned}
    \end{equation} 
    The integration along z-direction halts at the sample thickness $z_0$, and it has an analytical expression,
    \begin{equation}
    \begin{aligned}
    \int_0^{z_0} \text{FT}_{\text{2D}}\{h(\mathbf{r_{||}},z) \}(\mathbf{u})\text{d}z &= \frac{f}{k_0}\int_0^{z_0}   e^{-( \frac{\beta}{\rho_z}+ \frac{\langle \tan{\Delta \theta} \rangle \rho_l k_0^2 u^2}{2f^2} ) z}\text{d}z  \\
    &=  \frac{f}{k_0}\frac{1-e^{-( \frac{\beta}{\rho_z}+ \frac{\langle \tan{\Delta \theta} \rangle \rho_l k_0^2 u^2}{2f^2})z_0}   }{\frac{\beta}{\rho_z}+ \frac{\langle \tan{\Delta \theta} \rangle \rho_l k_0^2 u^2}{2f^2}}.
    \end{aligned}
    \end{equation}
    Finally, we obtain the explicit formula of the contribution from volumetric scattering,
    \begin{equation}\label{eq:FTs2}
    \begin{aligned}
    \text{FT}_{\text{2D}}\{ \langle s_{\text{3D}}(\mathbf{r}'_{1||}) \rangle\}(\mathbf{u}
    ) = \gamma(n^2-1)^2 \epsilon   \frac{f}{k_0}\text{FT}_{\text{2D}}\{I_i\} \frac{1-e^{-( \frac{\beta}{\rho_z}+ \frac{\langle \tan{\Delta \theta} \rangle \rho_l k_0^2 u^2}{2f^2})z_0}   }{\frac{\beta}{\rho_z}+ \frac{\langle \tan{\Delta \theta} \rangle \rho_l k_0^2 u^2}{2f^2}}
    \end{aligned}
    \end{equation}

    \subsection{Superposition model}
    Bring equation (\ref{eq:FTs1}) and (\ref{eq:FTs2}) into (\ref{eq:decom_principle}), we obtain the final expression of the field mutual intensity,
    \begin{equation}\label{eq:mutual_final} 
    \begin{aligned}
    \langle \mu(\mathbf{u}) \rangle& = \text{FT}_{\text{2D}} \{\langle s_{\text{2D}}(\mathbf{r}'_{1||}) \rangle + \langle s_{\text{3D}}(\mathbf{r}'_{1||}) \rangle\}(\mathbf{u})\\
    &= (n^2-1)^2\frac{\pi f^2}{k_0^2} \text{FT}_{\text{2D}}\{ I_i \}  \left( C_{\text{2D}}(\mathbf{u},p)+\kappa C_{\text{3D}}(\mathbf{u},p,z_0) \right),
    \end{aligned}
    \end{equation} 
    where $C_{\text{2D}}$ and $C_{\text{3D}}$ are the normalized decorrelations with following expressions,
    \begin{equation}\label{eq:C2D}
    \begin{aligned}
    &C_{\text{2D}}(\mathbf{u},p) = 
    \frac{\int \rho^2 \text{Jinc}(\frac{\rho k_0 |\mathbf{u}|}{f})p(\rho)\text{d}\rho}{\int \rho^2 p(\rho)\text{d}\rho}
    \end{aligned}
    \end{equation} 
    \begin{equation}\label{eq:C3D}
    \begin{aligned}
    &C_{\text{3D}}(\mathbf{u},p,z_0) = 
    \frac{1-e^{-\left( 1+ a (p) \frac{|\mathbf{u}|^2}{f^2} \right) \chi(p,z_0)} } {(1-e^{-\chi (p,z_0)})(1+ a (p) \frac{|\mathbf{u}|^2}{f^2} )} 
    \end{aligned}
    \end{equation} 
    \begin{equation}\label{eq:kappa}
    \begin{aligned}
    \kappa(p,z_0)=\frac{k_0 \gamma \epsilon  \rho_z(p)(1-e^{-\chi (p,z_0)})}{\pi \beta f}
    \end{aligned}
    \end{equation} 
    where $\chi(p,z_0)=\frac{\beta z_0}{\rho_z(p)}$, $a =  \frac{k_0^2 \langle \tan{\Delta \theta} \rangle^2 \rho_z^2(p) }{2\beta }$. As a reminder, $\rho_z(p)$ is the mean particle size under the distribution $p(\rho)$ along z-direction, $\beta$ is the dissipation rate at each scattering and $\Delta \theta$ is the scattering deflection angle, $\gamma$ is the volume filling factor of the disordered medium and $\epsilon $ is the coherent length along z-direction. In practice, we treat $\gamma \epsilon$, $\beta$ and $\langle \tan{\Delta \theta} \rangle$ as three unknown parameters in the system and fit them with calibrated samples. 

    Compare equation.(\ref{eq:mutual_final}) to equation.(\ref{eq:real_ME_3}), the normalized decorrelation $ C$ is, 
    \begin{equation}\label{eq:C}
    \begin{aligned}
    C = \frac{1}{\kappa+1}\left( C_{\text{2D}}+ \kappa  C_{\text{3D}}\right),
    \end{aligned}
    \end{equation} 
    and its normalization factor is $(n^2-1)^2\frac{\pi f^2}{k_0^2}(1+\kappa(p,z_0))$, indicating the total energy of the back-scattered light. 
    
    \section{Parameter fitting of the superposition model}
\begin{figure}[htb]
    \centering
    \includegraphics[width=0.6\linewidth]{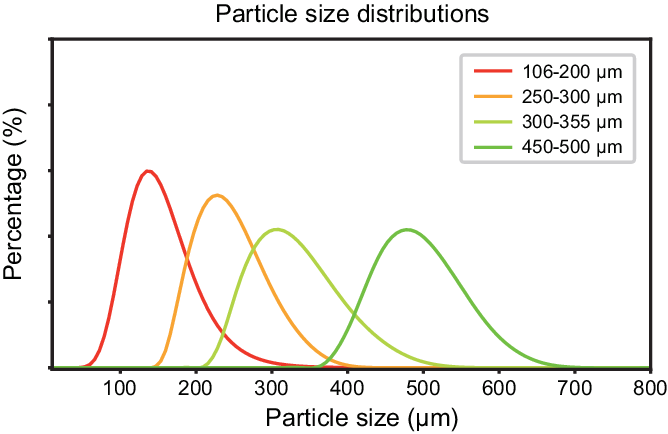}
    \caption{ Calibrated particle size distributions for sample sets in Fig.2.}
    \label{fig:s4}
    \end{figure}

\begin{figure}[htb]
    \centering
    \includegraphics[width=0.5\linewidth]{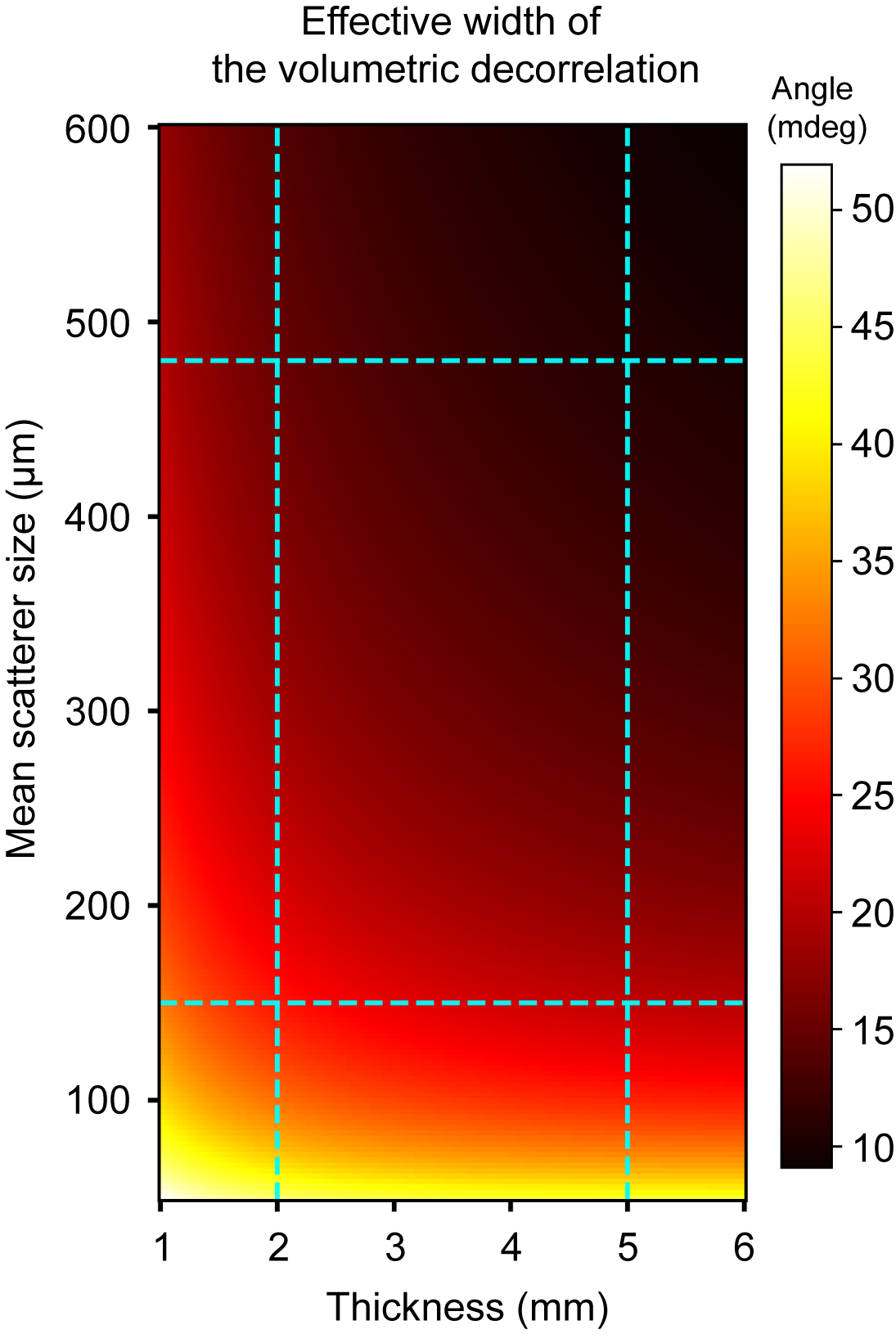}
    \caption{Map of the effective width $u_{\text{eff}}$ in $C_{\text{3D}}$.}
    \label{fig:s5}
    \end{figure}

    In Fig.2b in the main text, we show that the superposition model fits 16 different experimental images by only three parameters, achieving an over all L1-loss of less than 0.002. Four particle size distributions calibrated by a commercial offline particle size detector "Mastersizer" are presented in Fig.\ref{fig:s4}.
    The fitted values of $\gamma \epsilon$, $\beta$ and $\langle \tan \Delta \theta \rangle$ are 67.5 $\mu$m, 0.0257 and 3.10, respectively. 
    Fig.\ref{fig:s5} shows the effective width $u_{e}$ of $C_{\text{3D}}$ as a function of the thickness and mean scatterer size, based on the fitted parameters. The effective width is defined by the area underneath the decorrelation curve, $\pi u_{\text{eff}}^2=\int C_{\text{3D}}(\mathbf{u})\text{d}^2\mathbf{u}$. Dashed cyan lines indicate positions of curves shown in Fig.3(a), (b) and (c).
    
\begin{figure}[htb]
        \centering
        \includegraphics[width=0.5\linewidth]{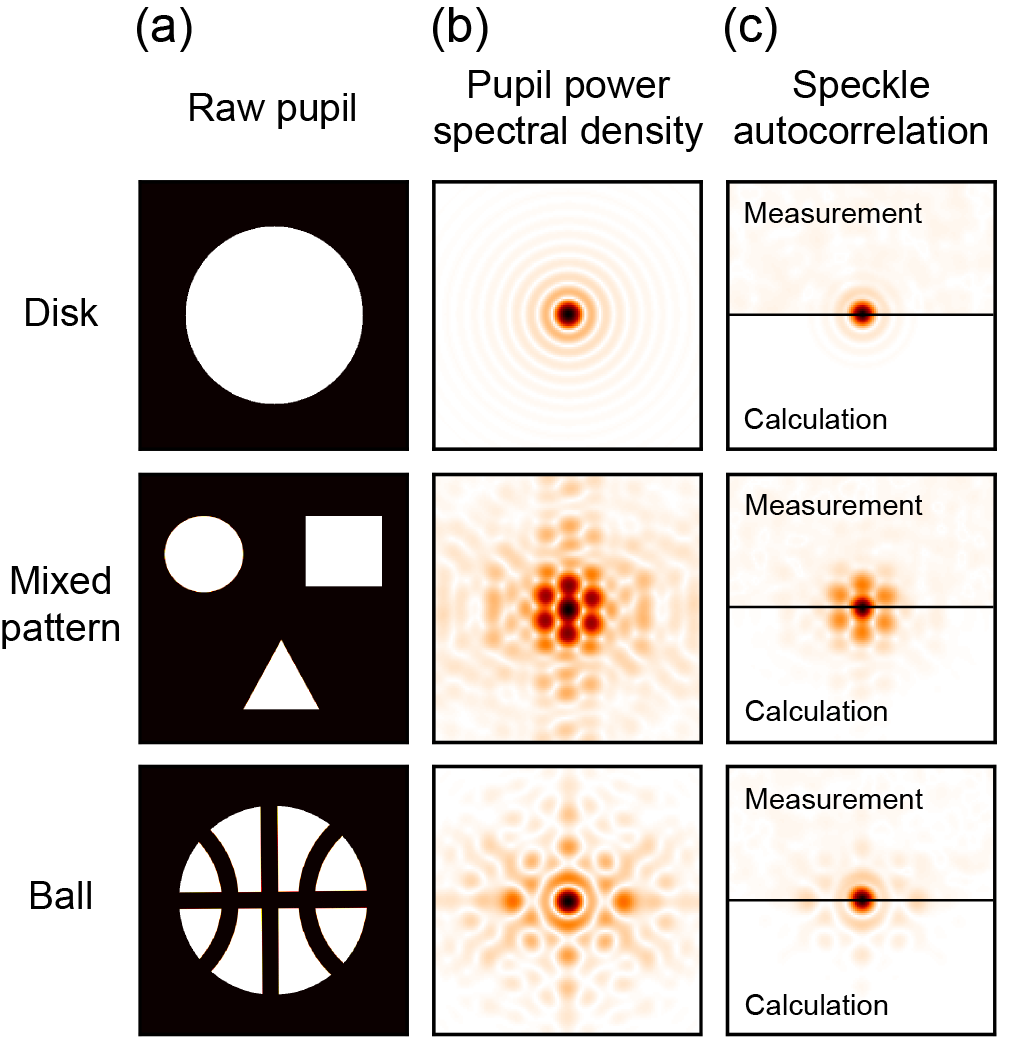}
        \caption{Scattering data with different incident pupil patterns. (a) Patterned pupils to shape the incident beam profile. (b) The corresponding power spectral densities of pupils. (c) Speckle autocorrelations measured from a 250-300 $\mu$m size and 2 mm thickness powder sample (upper panel) and the corresponding calculation results from the superposition model (lower panel).}
        \label{fig:s6}
    \end{figure}

    Under the extreme condition $z_0 \gg |\pmb{\rho}|$, $u_{e}$ follows the power law \cite{Feng1988} and is proportional to $\lambda f/l$, where the mean free path $l$ is equivalent to the mean scatterer size $\left<|\pmb{\rho}|\right>$ in this paper.
    Fig.\ref{fig:s6} presents more calculation results from the fitted superposition model for more incident patterns besides the cat shape.

\begin{figure}[htb]
        \centering
        \includegraphics[width=0.5\linewidth]{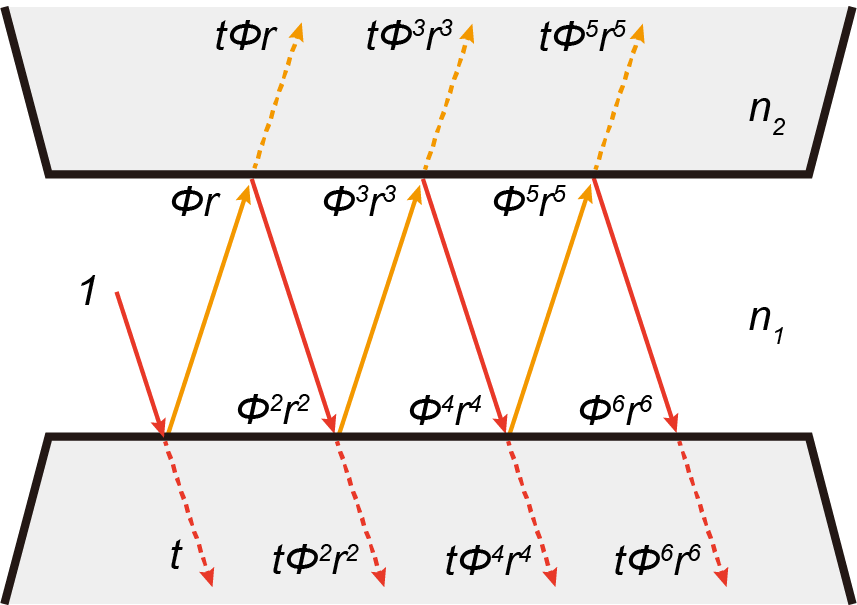}
        \caption{ A simplified model for backscattered energy flow $\beta$ between two powder surfaces. $n_1$ and $n_2$ denotes to the refractive index of vacuum and powders, respectively.  }
        \label{fig:s7}
    \end{figure}

    As our sample, KCl powder, has no absorption at 660 nm wavelength, the loss of the transmitted energy purely comes from the backward light in each scattering event. Therefore, $\beta$ can be estimated from a simplified model shown in Fig.\ref{fig:s7}, optical reflection and transmission between two smooth layers. 
    
    Fig.\ref{fig:s7} presents a local area in the powder surface, which is small enough to ensure the smoothness and large enough to ignore the wavefront shape. 
    The light propagates back and forth within two layers. For simplicity, we assume the incident direction is normal to the surface, so that the overall field is a superposition of different paths, although they are separated in the sketch for a better visibility. 

    The initial electrical field in the gap is $1$, $r$ and $t$ denote the field reflection and transmission ratio on the surface, $\Phi=e^{i\phi}$ is the phase shift induced by the free space propagation. Since the distance between particles is a random variable, $\Phi$ quickly rotate with a unity radius, eliminating the interference term. So the total backward energy flow between two surfaces is, 
    \begin{equation}
    \begin{aligned}
    \beta = |r|^2 -|r|^4+|r|^6-|r|^8 =\frac{ |r|^2 }{1+|r|^2},
    \end{aligned}
    \end{equation} 
    where $r=(n_2-n_1)/(n_2+n_1)$. For KCl powder, $n_2=1.5$, the calculated $\beta_{cal}=0.038$. The model to estimate $\beta$ is definitely over simplified, but the calculated result is the same order of magnitude with the measured one, validating the physical picture of $\beta$. 
    
    Although the refractive index $n$ only explicitly appears in the normalization factor from equation.(\ref{eq:C2D}) to equation.(\ref{eq:C}), it has an influence on the dissipation rate $\beta$. Additionally, according to Snell's law, the deflection angle and its tangent $\langle \text{tan} \Delta\theta \rangle$ also depend on $n$, but it is challenging to get an analytical expression, due to the complex scattering process. How these unknown parameters behave for different scattering materials is still an open-question, we will have a systematic study in future.

    \section{Background fluctuations in a single realization} 
    
    In this section, we analyze the fluctuations $\delta \mu$ in a single realization of the speckle autocorrelation. According to equation.(\ref{eq:single_mu}) and (\ref{eq:single_auto}), the standard deviation of the fluctuations in autocorrelation images is, 
    \begin{equation}\label{eq:single_autostd}
    \begin{aligned}
    \sqrt{\langle |\mu|^4 \rangle - \langle |\mu|^2 \rangle^2}& = \sqrt{\langle |\delta \mu|^4 \rangle - \langle |\delta \mu|^2 \rangle^2 + 4 |\langle \mu \rangle|^2\langle |\delta \mu|^2 \rangle}.
    \end{aligned}
    \end{equation}
    Since $\delta \mu \ll \langle \mu \rangle$, the first two terms in the right side of equation.(\ref{eq:single_autostd}) can be ignored. 
    The normalized standard deviation measured in Fig.4a of the main text is, 
    \begin{equation}\label{eq:single_normautostd}
    \begin{aligned}
    \frac{\sqrt{\langle |\mu|^4 \rangle - \langle |\mu|^2 \rangle^2}}{2|\langle \mu \rangle|^2 }& = \frac{\sqrt{ \langle |\delta \mu|^2 \rangle}}{|\langle \mu \rangle|}.
    \end{aligned}
    \end{equation}
    
    Following the same strategy for equations.(\ref{eq:u_s})(\ref{eq:s_2D_inter}) and (\ref{eq:s_2_1}), the single realization of the mutual intensity $\mu$ as,  
    \begin{equation}\label{eq:mutual_intensity_bcg}
    \begin{aligned}
     \mu(\mathbf{u}) = \text{FT}_{\text{2D}} \{ s_{\text{2D}} + s_{\text{3D}}  \}(\mathbf{u}),
    \end{aligned}
    \end{equation} 
    \begin{equation}\label{eq:s_peace_bcg}
    \begin{aligned}
     s_{\text{2D}} =  \frac{\pi f}{k_0} I_i(\mathbf{r}'_{||}) |\langle V (\mathbf{r}'_{||})\rangle_z |^2 ,
    \end{aligned}
    \end{equation}
    \begin{equation}\label{eq:s_diff_bcg}
    \begin{aligned}
    s_{\text{3D}} =   \epsilon \int V^2(\mathbf{r}'_{||},r'_{z})I(\mathbf{r}'_{||},r'_{z})  \text{d}r'_{1z},
    \end{aligned}
    \end{equation}
    
    where $\langle \cdot \rangle_z$ represents the spatial average along the z-direction over the top thin-layer with a thickness of $\Delta=\sqrt{\pi f/k_0}$. The Fourier transform in equation.(\ref{eq:mutual_intensity_bcg}) suggests that the background fluctuation $\delta\mu(\mathbf{u})$ at the large displacements $\mathbf{u}$ is corresponding to the relative high-frequency component in $s_{\text{2D}}$ and $s_{\text{3D}}$. 
        
    \begin{figure}[htb]
        \centering
        \includegraphics[width=1\linewidth]{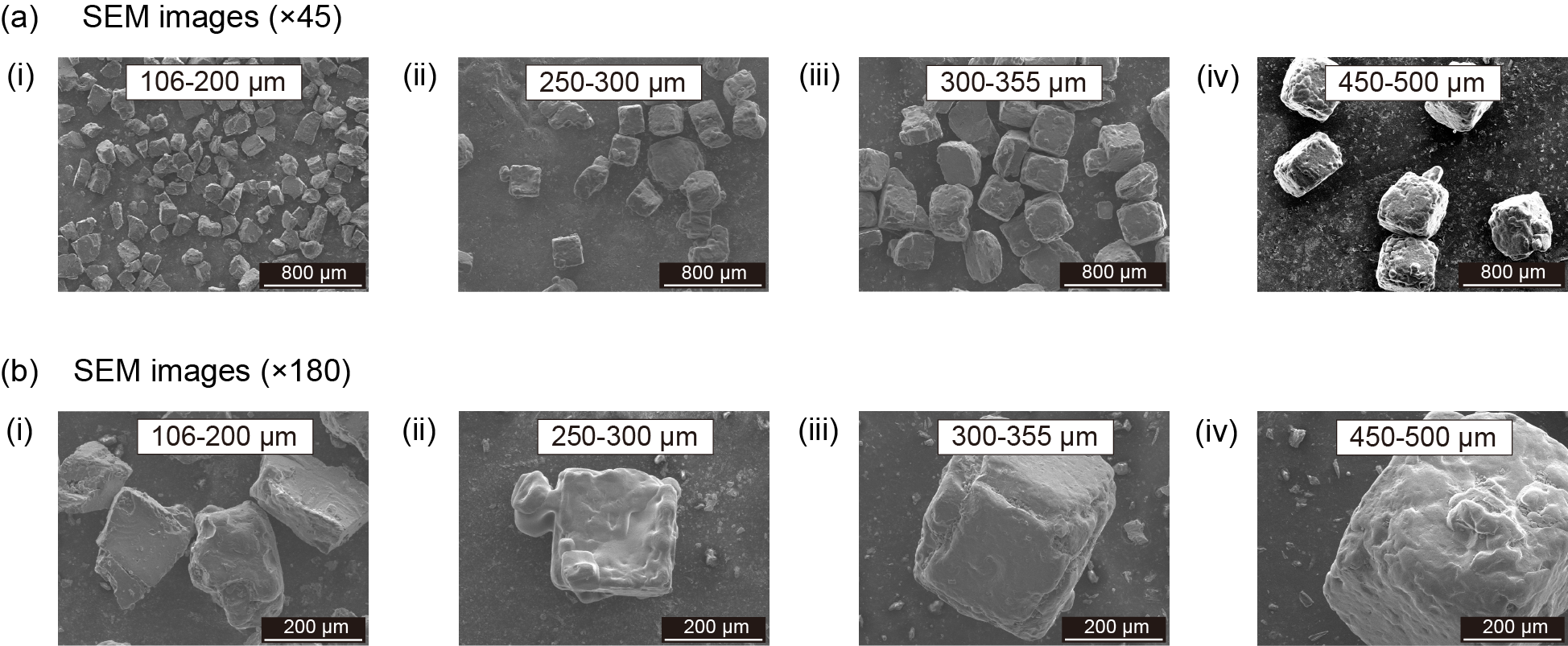}
        \caption{ Scanning electron microscopy (SEM) images for KCl powders. (a) $\times$ 45 magnification with a 800 $\mu$m scale bar; (b) $\times$ 180 magnification with a 200 $\mu$m scale-bar.
        }
        \label{fig:s9}
    \end{figure}
    
    Due to the interference inside the scattering medium, the volumetric intensity distribution $I$ incorporates a substantial amount of lateral high frequencies. The integral over the full sample space in $s_{\text{3D}}$ does not exhibit any size dependence, thereby causing the fluctuations of volumetric scattering to remain constant, denoted as $c$.
    
\begin{figure}[htb]
        \centering
        \includegraphics[width=0.6\linewidth]{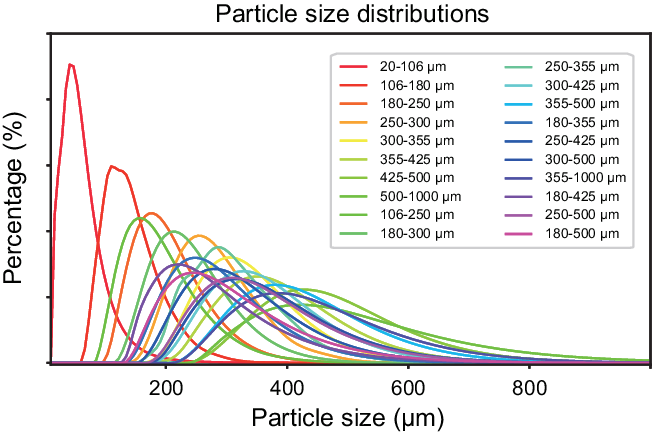}
        \caption{ Calibrated particle size distributions for sample sets in Fig.4.}
        \label{fig:s8}
    \end{figure}

    In $s_{\text{2D}}$, $I_i$ represents the incident beam profile, which is normally oriented to the sample surface, yielding a relatively smooth feature along the lateral direction. As illustrated in Fig.\ref{fig:s9} and Fig.4(c) in the main text, KCl particles have a cubic crystal structure with random scratches on the surface, and these scratches contribute to the high frequency component.     
    For simplification, we assume that the scratch depth $a$ is independent to particle size. The number of particles $N_0$ within the top layer $\Delta$ for a size $\rho$ is $N_0 = [\Delta/\rho]$, where $[\cdot]$ is the floor function. The surface high-frequency components in $\langle V (\mathbf{r}'_{||})\rangle_z$ are as follows, 
    \begin{equation}\label{eq:v_diff_bcg}
    \begin{aligned}
    \langle V (\mathbf{r}'_{||})\rangle_z = a \left(\xi_1(\mathbf{r}'_{||}) + \xi_2(\mathbf{r}'_{||}) + ... + \xi_{N_0}(\mathbf{r}'_{||})\right),
    \end{aligned}
    \end{equation}
    where $\xi_1$, $\xi_2$,..., $\xi_{N_0}$ represent $N_0$ distinct normalized white noise functions. The equivalent amplitude of $|\langle V (\mathbf{r}'_{||})\rangle_z|^2$ is $N_0 a^2$. After the normalization, the contribution from $s_{\text{2D}}$ to the normalized standard deviation of autocorrelations is proportional to $N_0a^2/ \Delta^2$.
    Based on the above discussion, we derive that the background fluctuation is given by the following equation, 

    \begin{equation}\label{eq:bcg}
    \begin{aligned}
    \frac{\sqrt{ \langle |\delta \mu|^2 \rangle}}{|\langle \mu \rangle|} &= c + \frac{a^2}{\Delta^2} \int N_0(\rho) p_v(\rho)\text{d}\rho\\
    &= c + \frac{a^2}{\Delta^2} \int \left[ \frac{\Delta}{\rho} \right] p_v(\rho)\text{d}\rho,
    \end{aligned}
    \end{equation} 
    where $p_v(\rho)$ is the volume-based particle size distribution, defines as $p_v(\rho)=\frac{p(\rho)\rho^3}{\int p(\rho)\rho^3 \text{d}\rho}$, $p(\rho)$ is the number-based particle size distribution. $a$, $\Delta$ and $c$ are three unknown parameters in the system, which are fitted using 20 samples with different distributions as depicted in Fig.\ref{fig:s8}.

    \section{Retrieval algorithm}
    In this section, we derive the image retrieval algorithm with the perspective of optimization.

    The inversion of the forward model of the amnesia effect presents an optimization problem:
    \begin{equation}
    E(O)=\frac{1}{2}\left\Vert\left| \text{FT}_{2D}\{ |O|^2 \} \times P \right|-\sqrt{M}\right\Vert_{2}^{2}\label{eq:optim_obj}
    \end{equation}
    \begin{equation}\label{eq:optim_main}
    \widehat{O}=\mathop{\arg\min}\limits_{O} E(O)
    \end{equation}
    Equation.(\ref{eq:optim_obj}) gives the objective function defined based on the forward model of the amnesia effect, where $M=\left \langle I \star I \right \rangle$ is the mean autocorrelation of speckle measurement intensity.
    
    Subsituting the Fourier transform of the object intensity profile with an intermediate variable $\Psi=\text{FT}_{2D}\{ |O|^2 \}$, then equation.(\ref{eq:optim_main}) can be solved with the two-step retrieval:

    \begin{subequations}
    \begin{align}
    \begin{split}
    \widehat{\Psi}&=\mathop{\arg\min}\limits_{\Psi} E(\Psi) \\ 
    &=\mathop{\arg\min}\limits_{\Psi} \frac{1}{2}\left\Vert\left| \Psi P \right|-\sqrt{M}\right\Vert_{2}^{2}\label{eq:optim1}
    \end{split}\\
    \begin{split}
    |\widehat{O}| &= \sqrt{ Proj_{\mathbb{R}}\{\text{IFT}_{2D}\{ \widehat{\Psi}\}  \} }\label{eq:optim2}
    \end{split}
    \end{align}
    \end{subequations}
    where $\widehat{\Psi}$ and $|\widehat{O}|$ represent the optimal solution of the introduced intermediate variable and the pupil modulus profile, repectively. Equation.(\ref{eq:optim2}) describes a quite understandable projection operation, which has been commonly utilized in correlation imaging \cite{Bertolotti2012,Katz2014}. $Proj_{\mathbb{R}}$ projects a complex vector to the real domain, which is the constraint given by the forward model.

    Noting that in equation.(\ref{eq:optim1}) E is a real-valued function with complex arguments, it can be studied with Wirtinger calculus. By calculating the Wirtinger gradient of $E(\Psi)$, the optimization problem can be solved with gradient descent method.
    \begin{equation}\label{eq:gradient}
    \begin{aligned}
    \nabla E(\Psi) &= \frac{\partial E}{\partial \Psi^*} \\
    &= \frac{\partial \sqrt{\Psi P (\Psi P)^*}}{\partial \Psi^*} (\sqrt{|\Psi P|^2} - \sqrt{M}) \\
    &= \frac{\Psi P^2}{2} \left(1-\frac{\sqrt{M}}{\sqrt{|\Psi P|^2}}\right) 
    \end{aligned}
    \end{equation}

    Gradient descent gives the update function in each iteration:
    \begin{equation}\label{eq:GD}
    \Psi'=\Psi-\beta\nabla E_{\Psi}
    \end{equation}
    where $\Psi'$ represents the updated intermediate variable and $\beta$ is the step size.

    Substitute Eq.\eqref{eq:GD} into Eq.\eqref{eq:optim_obj} helps find the optimal step size:
    \begin{equation}\label{eq:update}
    \begin{aligned}
    E(\Psi, \beta)&=\frac{1}{2}\left\Vert\left| (\Psi-\beta\nabla E(\Psi)) P \right|-\sqrt{M}\right\Vert_{2}^{2} \\
    &=\frac{1}{2}\left\Vert (\left| \Psi P \right|-\sqrt{M})(1-\frac{\beta}{2}P^2) \right\Vert_{2}^{2}
    \end{aligned}
    \end{equation}
    Finding the optimal $\beta$ to minimize the objective function $E$ in equation.\eqref{eq:update} corresponds to a one-dimensional least-square optimization since $\beta$ is a real number and all other elements are known vectors. The optimal step size is given by:
    \begin{equation}\label{eq:step_size}
    \widehat\beta=\frac{2\left\Vert (|\Psi P|-\sqrt{M})P \right\Vert_{2}^{2}}{\left\Vert (|\Psi P|-\sqrt{M})P^2 \right\Vert_{2}^{2}}
    \end{equation}
    Note that the filter $P$ here is a real vector in the decorrelation, while in previous occasion \cite{Bertolotti2012,Katz2014} $P=1$, the optimal step size is set as $\widehat\beta=2$. This corresponds to the well-known modulus projection (or error reduction algorithm) brought up by Fienup \cite{Fienup1982}.
\end{document}